\documentclass{aastex631}
\usepackage{natbib}
\usepackage{graphicx}
\usepackage[flushleft]{threeparttable}
\usepackage{mathtools}
\usepackage{xcolor}
\usepackage{amsmath} 
\usepackage{enumerate}
\usepackage{enumitem}
\usepackage[normalem]{ulem} 
\usepackage{bm}

\shorttitle{reconstructed mass map in merging cluster}
\shortauthors{Chen et al.}

\begin{document}
\title{Mapping Dark Matter in Merging Clusters with Weak Lensing for CSST: Morphology and Filament Recovery}
\author[0009-0009-8953-1839]{Xinyue Chen}
\footnote{chenxinyue@bao.ac.cn}
\affiliation{National Astronomical Observatories (NAOC), Chinese Academy of Sciences, Beijing 100101, China}
\affiliation{School of Astronomy and Space Science, University of Chinese Academy of Science, Beijing 100049, China}

\author[0000-0003-3899-0612]{Ran Li}\footnote{liran@bnu.edu.cn}
\affiliation{School of Physics and Astronomy, Beijing Normal University,  Beijing 100875, China}
\affiliation{School of Astronomy and Space Science, University of Chinese Academy of Science, Beijing 100049, China}

\author{Qiao Wang}
\affiliation{National Astronomical Observatories (NAOC), Chinese Academy of Sciences, Beijing 100101, China}
\affiliation{School of Astronomy and Space Science, University of Chinese Academy of Science, Beijing 100049, China}

\author{Hengjie Lin}
\affiliation{National Astronomical Observatories (NAOC), Chinese Academy of Sciences, Beijing 100101, China}

\author{Liang Gao}
\affiliation{School of Physics and Astronomy, Beijing Normal University,  Beijing 100875, China}
\affiliation{School of Astronomy and Space Science, University of Chinese Academy of Science, Beijing 100049, China}

\author{Xiangkun Liu}
\affiliation{ South-Western Institute for Astronomy Research (SWIFAR), Yunnan University, Kunming 650500, PR China}
\affiliation{Yunnan Key Laboratory of Survey Science,Yunnan University, Kunming 650500, PR China}

\author[0000-0001-8534-837X]{Huanyuan Shan}
\affiliation{Shanghai Astronomical Observatory (SHAO), Nandan Road 80, Shanghai 200030, China}
\affiliation{Key Laboratory of Radio Astronomy and Technology, Chinese Academy of Sciences, A20 Datun Road, Chaoyang District, Beijing, 100101, P. R. China}
\affiliation{University of Chinese Academy of Sciences, Beijing 100049, China}

\author[0000-0002-7336-2796]{Ji Yao}
\affiliation{Shanghai Astronomical Observatory (SHAO), Nandan Road 80, Shanghai 200030, China}
\affiliation{University of Chinese Academy of Sciences, Beijing 100049, China}

\author{Ming Li}
\affiliation{National Astronomical Observatories (NAOC), Chinese Academy of Sciences, Beijing 100101, China}
\affiliation{School of Astronomy and Space Science, University of Chinese Academy of Science, Beijing 100049, China}

\begin{abstract}

Weak gravitational lensing offers a direct probe of the projected dark matter distribution in cluster-scale structures and the surrounding cosmic web. We investigate the capability of the China Space Station Telescope (CSST) to recover non-spherical dark matter morphologies using mock reduced-shear maps generated from cosmological N-body simulations. We apply the deep-learning-based MIU$^{2}$Net mass-reconstruction framework to CSST-like wide- and deep-field source densities, focusing on two representative systems: filament-connected halo pairs and elongated halos. MIU$^{2}$Net accurately recovers halo centers, elongated mass distributions, and the filamentary structures connecting halo pairs in both survey configurations. Halo centers are localized within about 0.1 $\,h^{-1}\,\mathrm{Mpc}$, and elongated-halo shapes and orientations are reliably recovered. Filament paths are stable against shape noise, with mean normalized instability parameters $\mathcal{I}_{F}$, defined as the mean deviation from the true filament locus in units of $R_{200}$,
decreasing from 0.041 to 0.037 in the low-redshift bin and from 0.051 to 0.045 in the high-redshift bin from the wide to deep field. The recovered filament shapes remain consistent with the true maps at the percent level. Compared with conventional Kaiser-Squires reconstruction, MIU$^{2}$Net better preserves low-contrast anisotropic structures while suppressing small-scale noise. These results demonstrate that CSST weak-lensing data, combined with machine-learning-based mass reconstruction, will enable statistical studies of cluster assembly, anisotropic matter accretion, and the connection between massive halos and the filamentary cosmic web.

\keywords{
gravitational lensing: weak --
methods: statistical --
dark matter --
large-scale structure of Universe
}

\end{abstract}

\section{Introduction}

Dark matter (DM) dominates the matter content of the Universe and governs the formation and evolution of large-scale structure (LSS), including clusters, filaments, walls, and voids, through gravitational instability and hierarchical matter assembly \citep{1984Natur.311..517B,1985ApJ...292..371D,1996Natur.380..603B,2005Natur.435..629S,1980lssu.book.....P}. Accurately mapping the spatial distribution of DM is therefore crucial for understanding how massive halos assemble within the cosmic web and for testing the physical nature of DM itself. In the cosmic-web picture, massive halos preferentially form at the nodes of the LSS and grow through anisotropic accretion and mergers along filaments. Simulations have shown that cluster formation is closely linked to the surrounding filamentary environment, with matter infall occurring along a few preferred directions that correspond to the locations of filaments at later times \citep{1996Natur.380..603B,2005MNRAS.359..272C}. Halo pairs connected by filaments and elongated halos are therefore expected signatures of anisotropic mass assembly, encoding information about the tidal field, merger history, and connectivity of the cosmic web \citep{dietrich2012filament,2014MNRAS.443.1090F,2021PhRvD.103f3517H}. Moreover, merging halo systems provide important laboratories for testing DM physics: offsets between galaxies, hot gas, and the lensing-inferred mass distribution, as observed in systems such as the Bullet Cluster, have been used to constrain whether DM is effectively collisionless or self-interacting \citep{Markevitch_2004,2006ApJ...648L.109C,2008ApJ...679.1173R}. Reliable reconstruction of halo pairs, elongated halos, and their connecting filaments is therefore essential for both cosmic-web studies and tests of DM physics.



Weak gravitational lensing (WL) provides a direct observational probe of the projected matter distribution in the Universe. By measuring the coherent distortions, or shear, of background galaxy images, WL enables the reconstruction of the dimensionless surface mass density, or convergence $\kappa$, map, without relying on assumptions about the dynamical state or baryonic composition of the lensing system \citep{bartelmann2001weak,2015MNRAS.450.2888L,Liu_2019,Tam_2020}. This makes WL particularly powerful for merging clusters and massive halo-pair systems, where the distributions of galaxies, hot gas, and dark matter can be spatially offset and where the mass morphology is often highly asymmetric and multi-peaked.

However, recovering the detailed morphology of merging clusters and their connecting filaments remains challenging. The WL signals of filaments and elongated halos are typically low contrast and spatially extended, and their reconstruction is limited by the finite source-galaxy number density, shape noise, and projection effects. Space-based imaging offers a major advantage for this problem by providing a stable point-spread function, high angular resolution, and a higher density of resolved background galaxies. The China Space Station Telescope (CSST) will conduct a wide-field photometric survey with high-resolution imaging, reaching a depth of $\sim$25.5 mag in multiple bands, an angular resolution of $\sim$ 0.15-0.2 arcsec, and a planned sky coverage of $\sim$17,500 deg$^2$ \citep{2026SCPMA..6939501C}. These characteristics make CSST especially well suited for WL studies of complex cluster-scale dark matter structures, while its wide sky coverage will provide a large statistical sample of such systems.


Nevertheless, translating CSST-like shear measurements into reliable convergence maps remains non-trivial. 
Early efforts to extract clusters from noisy shear data included the
tomographic matched filter of \citet{Hennawi_2005}, which incorporates
source photometric-redshift information, and the optimal linear filter
of \citet{Maturi_2005}, which accounts for both intrinsic shape noise
and projected large-scale-structure noise. These methods were designed
primarily for cluster detection. For direct convergence-map
reconstruction, widely used approaches include the Kaiser--Squires
(KS) inversion \citep{1993ApJ...404..441K}, Wiener filtering
\citep{2017ApJ...839...25S,Jeffrey_2018}, and the wavelet-based
multiscale-entropy method of \citet{Starck_2006}. However, KS inversion is sensitive to shape noise, Wiener filtering can oversmooth small-scale structures, and wavelet-based methods depend on the adopted regularization and detection thresholds.

Recent advances in machine learning (ML) provide promising alternatives for WL-based mass reconstruction \citep{2019PhRvD.100d3527S,2020MNRAS.492.5023J,2021ApJ...923..266H}. When trained and validated on realistic simulations, convolutional neural networks can learn non-linear mappings from noisy shear fields to the underlying convergence distribution, potentially preserving morphological information more effectively than linear inversion methods. In this work, we use the MIU$^{2}$Net framework proposed in \citet{2026RAA....26g5003H} to evaluate the recovery of dark matter morphology and filamentary structures in CSST-like observations. We construct mock shear maps from cosmological $N$-body simulations, adopt both wide- and deep-field source-galaxy number densities expected for CSST, and compare the reconstructed convergence maps with ground-truth. Our goal is to assess whether CSST-quality WL data, combined with ML-based reconstruction, can robustly recover halo nodes, elongated mass distributions, and the low-contrast filaments connecting merging cluster systems.

The structure of this paper is as follows. Section 2 reviews weak-lensing theory and simulations, Section 3 describes the methodology, Section 4 presents the results, and Section 5 summarizes our conclusions.

\begin{table*}
\centering
\caption{The bin criteria.}
\begin{threeparttable}
\small
\setlength{\tabcolsep}{4pt}
\begin{tabular}{c|ccccc}
\hline\hline
Sample & Type & $z$
& $\log_{10}[M/(h^{-1}M_\odot)]$
& Center Separation($h^{-1}\mathrm{Mpc}$)/Axis Ratio
& Number \\
\hline

Pair(z1m1d1) & Halo Pair & 0.1--0.2 & $14.5$--$14.8$ & $<5$ & 602 \\
Pair(z1m1d2) & Halo Pair & 0.1--0.2 & $14.5$--$14.8$ & 5--10 & 1197 \\
Pair(z1m2d1) & Halo Pair & 0.1--0.2 & $14.8$--$15.5$ & $<5$ & 412 \\
Pair(z1m2d2) & Halo Pair & 0.1--0.2 & $14.8$--$15.5$ & 5--10 & 1125 \\

Pair(z2m1d1) & Halo Pair & 0.2--0.3 & $14.5$--$14.8$ & $<5$ & 665 \\
Pair(z2m1d2) & Halo Pair & 0.2--0.3 & $14.5$--$14.8$ & 5--10 & 1487 \\
Pair(z2m2d1) & Halo Pair & 0.2--0.3 & $14.8$--$15.5$ & $<5$ & 505 \\
Pair(z2m2d2) & Halo Pair & 0.2--0.3 & $14.8$--$15.5$ & 5--10 & 1207 \\

Elongated(z1m1q1) & Elongated Halo & 0.1--0.2
& $15.0$--$15.2$ & $\leq q_{15,z1}$ & 14 \\
Elongated(z1m1q2) & Elongated Halo & 0.1--0.2
& $15.0$--$15.2$ & $>q_{15,z1}$ & 127 \\
Elongated(z1m2q1) & Elongated Halo & 0.1--0.2
& $15.2$--$15.5$ & $\leq q_{15,z1}$ & 26 \\
Elongated(z1m2q2) & Elongated Halo & 0.1--0.2
& $15.2$--$15.5$ & $>q_{15,z1}$ & 96 \\

Elongated(z2m1q1) & Elongated Halo & 0.2--0.3
& $15.0$--$15.2$ & $\leq q_{15,z2}$ & 18 \\
Elongated(z2m1q2) & Elongated Halo & 0.2--0.3
& $15.0$--$15.2$ & $>q_{15,z2}$ & 121 \\
Elongated(z2m2q1) & Elongated Halo & 0.2--0.3
& $15.2$--$15.5$ & $\leq q_{15,z2}$ & 15 \\
Elongated(z2m2q2) & Elongated Halo & 0.2--0.3
& $15.2$--$15.5$ & $>q_{15,z2}$ & 61 \\

\hline\hline
\end{tabular}

\begin{tablenotes}[flushleft]
\item \textbf{Note}:
The columns list, in order, the subsample name, type, redshift range,
main-halo mass range, projected center-separation or axis-ratio
($q=b/a$) criterion, and sample size. The quantities $q_{15,z1}$ and
$q_{15,z2}$ denote the 15th percentiles of the true projected
axis-ratio distributions in the $z1$ ($0.1<z\leq0.2$) and $z2$
($0.2<z\leq0.3$) bins, respectively, with
$q_{15,z1}=0.167$ and $q_{15,z2}=0.184$.
Thus, the $q1$ subsamples contain the most elongated 15\% of halos in
each redshift bin, while the $q2$ subsamples contain the remaining
85\%.
\end{tablenotes}

\end{threeparttable}
\label{tab:bin}
\end{table*}

\section{WEAK LENSING THEORY AND SIMULATIONS}
\label{sec:Theory}

\subsection{Formalism}
\label{sec:Formalism}

This section lays out the theoretical framework of weak gravitational lensing \citep{bartelmann2001weak,2018ARA&A..56..393M}. Weak lensing relies on the subtle distortions imprinted on background galaxy shapes by the gravity of intervening mass structures. The degree of this distortion is directly linked to the convergence field, $\kappa$, which in turn is determined by the matter overdensity, $\delta$:

\begin{equation}
\label{eq:kappa_delta_w}
\kappa(\boldsymbol{\theta}, w)
=
\frac{3 H_0^2 \Omega_{\rm m}}{2 c^2}
\int_0^{w_{\rm h}} dw \,
\frac{g(w)\, f_K(w)}{a(w)}\,
\delta\!\left(f_K(w)\boldsymbol{\theta}, w\right),
\end{equation}

with the lensing efficiency function
\begin{equation}
\label{eq:lensing_efficiency}
g(w)
=
\int_w^{w_{\rm h}} dw'\,
p_w(w')\,
\frac{f_K(w'-w)}{f_K(w')},
\end{equation}
where $w$ denotes the comoving distance and $p_w(w)$ connected to the redshift distribution $p_z(z)$ via
$p_z(z)\,dz = p_w(w)\,dw$.
Here $a(w)$ is the scale factor, and $f_K(w)$ represents the comoving angular
diameter distance.

The lensing effect is mathematically described by a scalar quantity known as the lensing potential, $\psi(\theta)$, from which all other lensing observables are derived. The deflection of light rays is given by the potential's gradient:
\begin{equation}
\alpha = \nabla\psi,
\end{equation}
while the convergence, which quantifies the local magnification of background sources, is directly linked to the potential's second derivatives via the 2D Poisson equation:
\begin{equation}
\label{eq:kappa_psi}
\kappa = \frac{1}{2}\nabla^2\psi = \frac{1}{2}\left(\frac{\partial^2}{\partial\theta_1^2} + \frac{\partial^2}{\partial\theta_2^2}\right)\psi,
\end{equation}
where the partial derivatives are taken with respect to the angular coordinates $\theta_1$ and $\theta_2$ on the sky.

The stretching or ``shear" of galaxy shapes is captured by the two-component shear, $\gamma = \gamma_1 + i\gamma_2$. Both components are also derived from the lensing potential:
\begin{equation}
\gamma_1 = \frac{1}{2}\left(\frac{\partial^2}{\partial\theta_1^2} - \frac{\partial^2}{\partial\theta_2^2}\right)\psi, \quad \gamma_2 = \frac{\partial^2}{\partial\theta_1\partial\theta_2}\psi.
\end{equation}
A fundamental relationship exists between the shear and convergence fields, known as the Kaiser-Squires (KS) transform \citep{1993ApJ...404..441K}. In real space, this takes the form of a convolution, while its Fourier space representation is more computationally efficient, converting the complex convolution into a simple multiplication:
\begin{equation}
\tilde{\gamma}_1(\bm{l}) = \frac{l_x^2 - l_y^2}{\bm{l}^2} \tilde{\kappa}(\bm{l}), \quad 
\tilde{\gamma}_2(\bm{l}) = \frac{2l_x l_y}{\bm{l}^2} \tilde{\kappa}(\bm{l}),
\end{equation}
where $\tilde{\gamma}$ and $\tilde{\kappa}$ are the Fourier transforms of the shear and convergence fields, respectively.

In a practical observational setting, a linear data model is often adopted to connect the theoretical convergence map to the noisy shear measurements:
\begin{equation}
\label{eq:data_model}
\bm{\gamma}_{\text{obs}} = \mathbf{A}\bm{\kappa} + \mathbf{n},
\end{equation}
where $\bm{\gamma}_{\text{obs}}$ and $\bm{\kappa}$ are vectors for shear and convergence on an image grid, $\mathbf{A}$ is a matrix operator, and $\bm{n}$ is the noise vector.

It is crucial to distinguish between the true shear, $\gamma$, and the directly measured quantity, the ``reduced shear", $g$, defined as:
\begin{equation}
g_{1,2} = \frac{\gamma_{1,2}}{1 - \kappa}.
\label{eq:reduced}
\end{equation}
In most weak-lensing regimes, where both $\gamma$ and $\kappa$ are small, the reduced shear provides an excellent approximation to the true shear, i.e., $g \approx \gamma$. However, this approximation becomes less accurate in regions of high mass concentration, such as cluster centers, where $\kappa$ is non-negligible and the distinction between shear and reduced shear becomes important \citep{2009ApJ...696..775S}. In this work, we use the reduced shear field as the model input and the corresponding convergence field as the target for training.

\begin{figure*}
\centering
\includegraphics[width=0.8\textwidth]{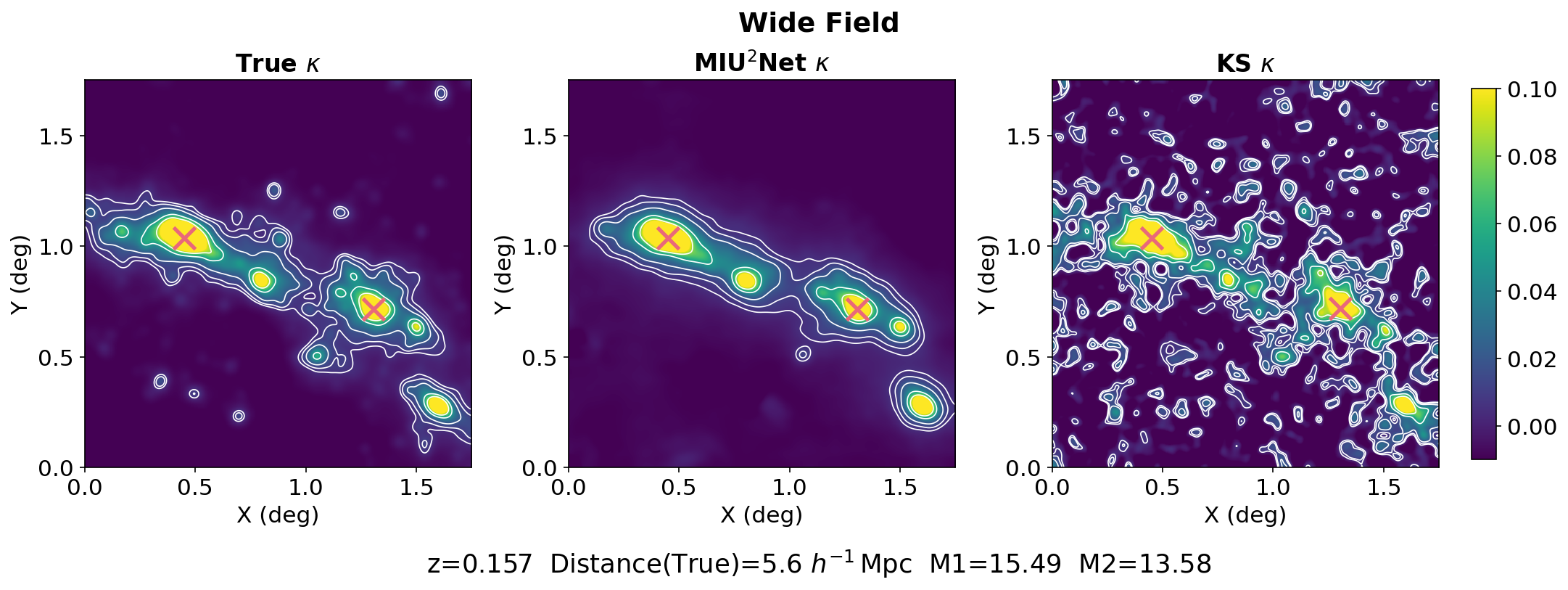}
\includegraphics[width=0.8\textwidth]{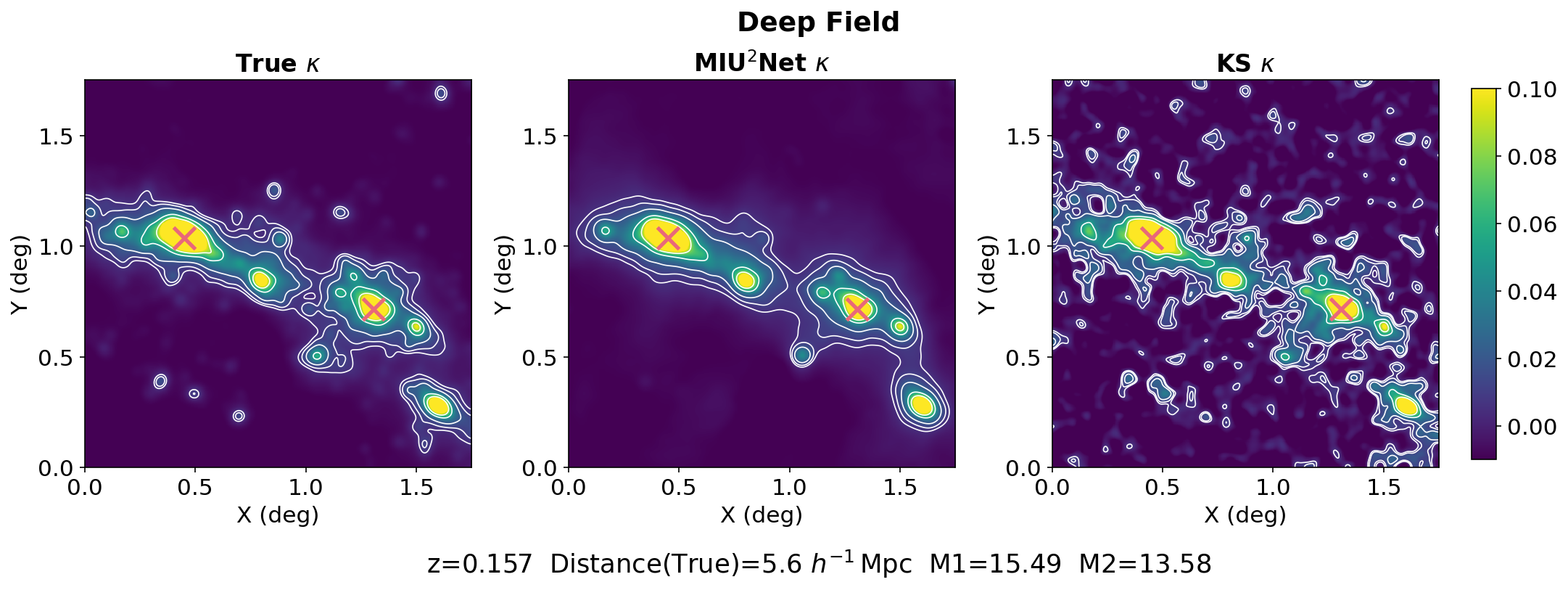}
\includegraphics[width=0.8\textwidth]{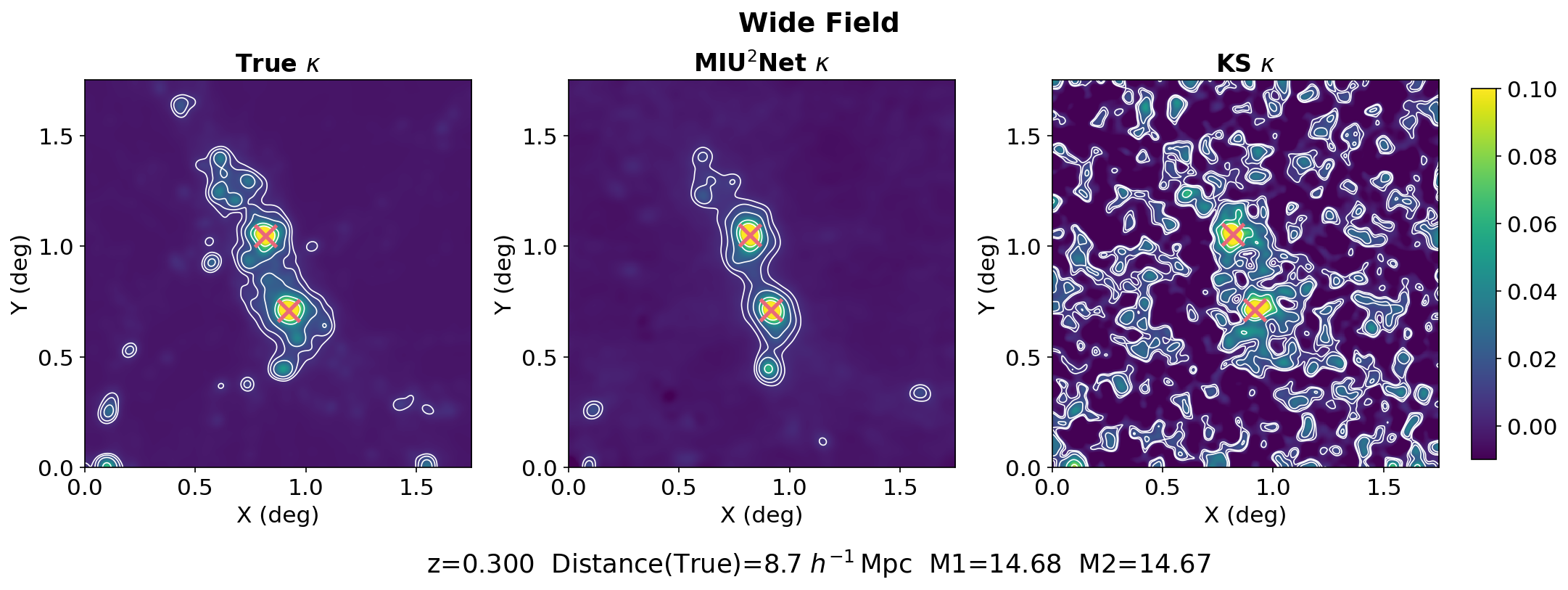}
\includegraphics[width=0.8\textwidth]{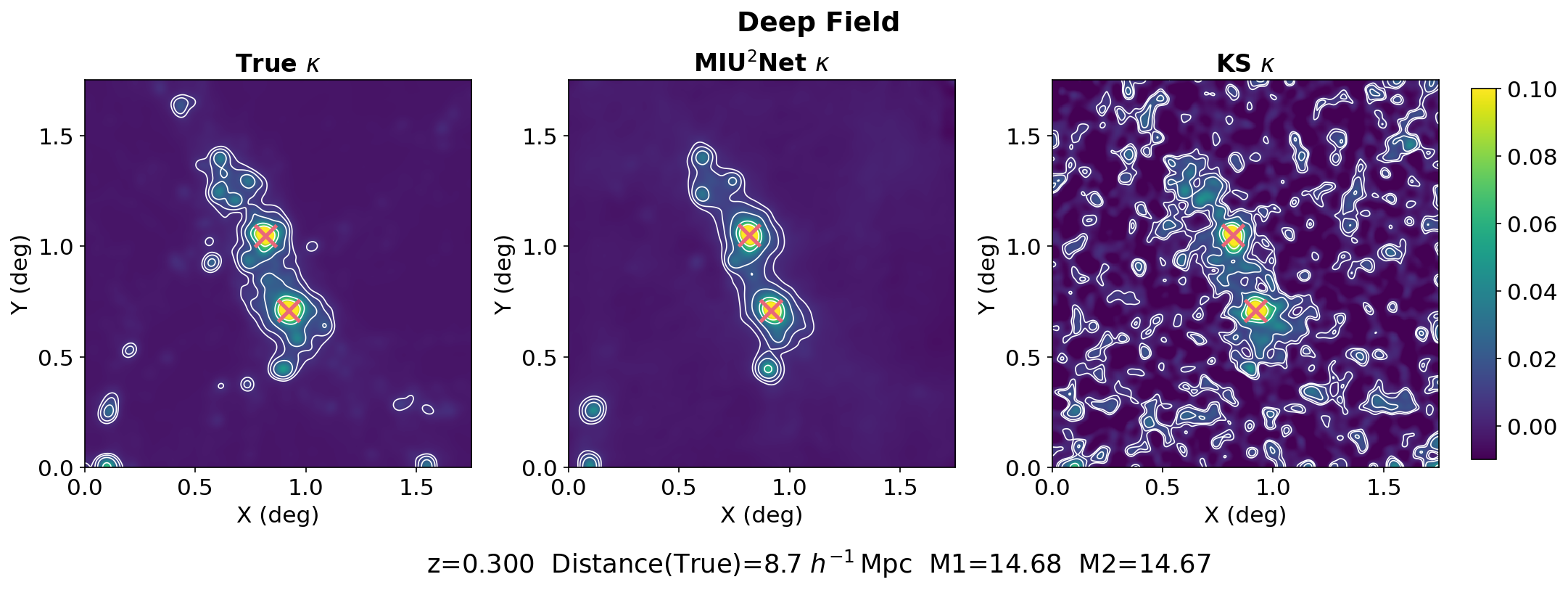}
 \caption{ Two example convergence maps for halo-pair systems. From left to right, the panels show the ground-truth, MIU$^{2}$Net, and Kaiser--Squires (KS)
maps, with the wide- and deep-field cases presented separately. All
maps have a pixel size of $0.21$ arcmin and are smoothed with a
1-arcmin Gaussian kernel. The contours at
$\kappa=[0.005,0.01,0.02,0.05,0.08]$ trace structures at different
convergence levels. The two peaks associated with each halo pair are
marked by crosses in all maps.
}
\label{fig:comparison1} 
\end{figure*}

\begin{figure*}
\centering
\includegraphics[width=0.48\textwidth]{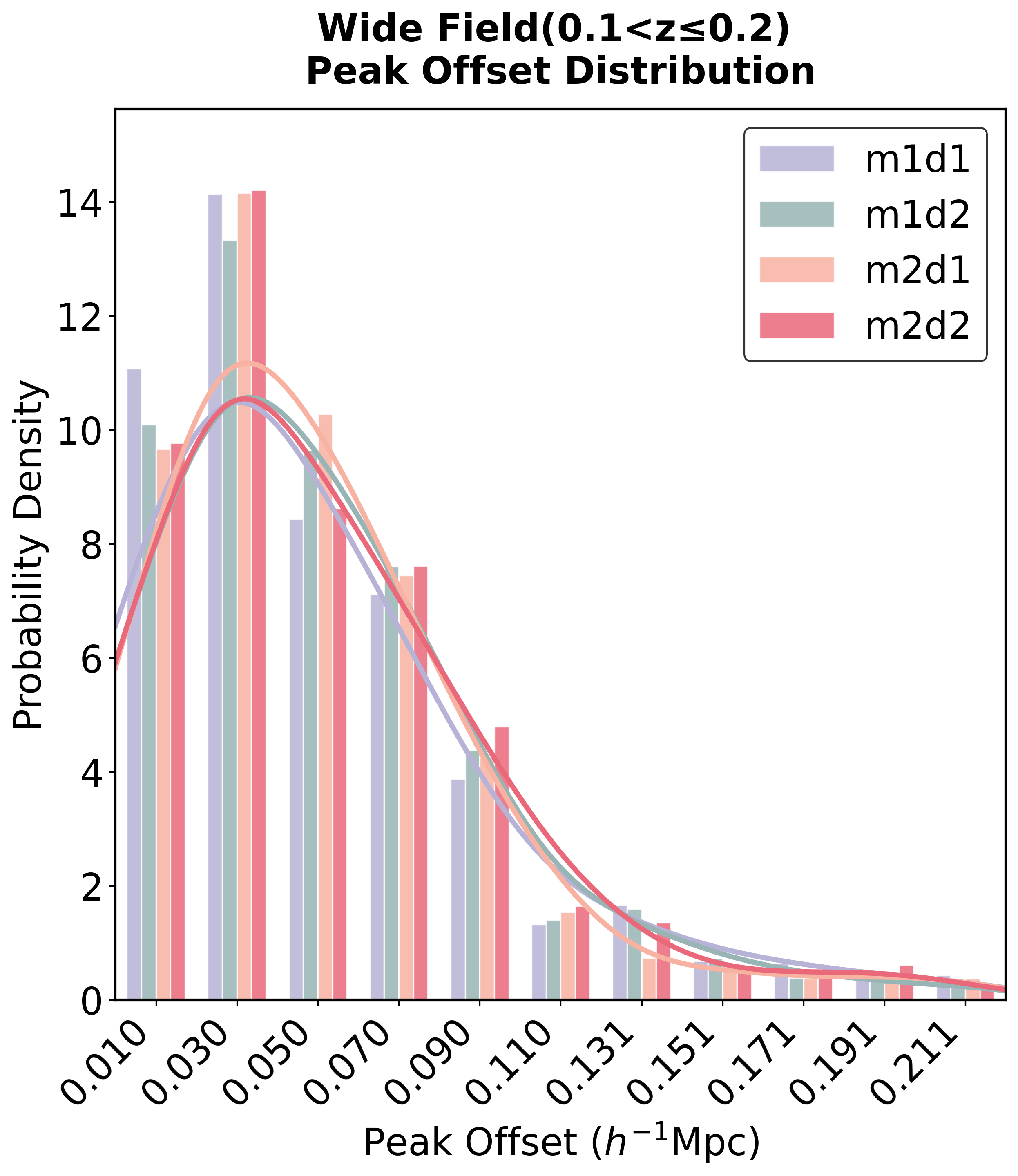}
\includegraphics[width=0.495\textwidth]{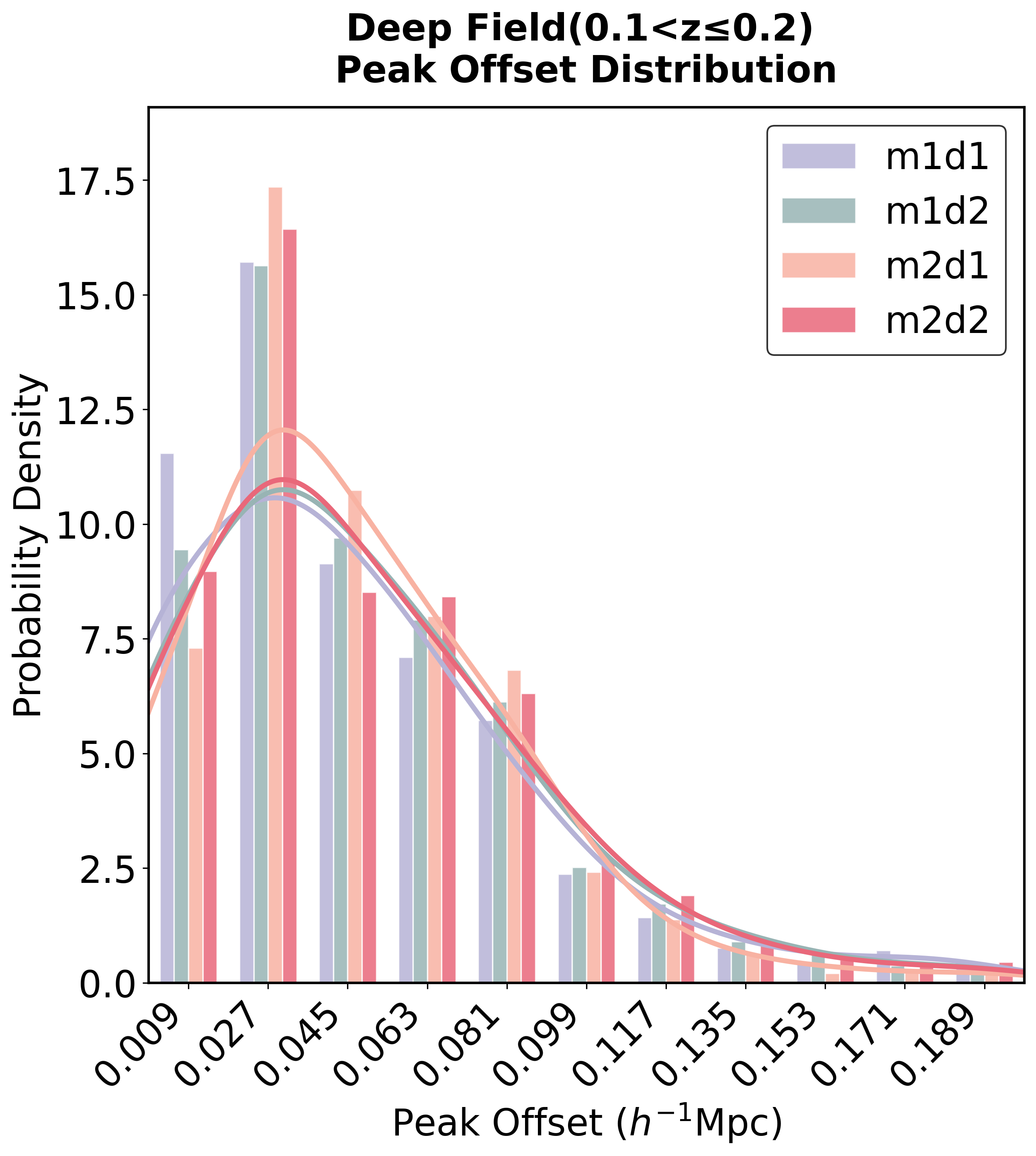}
\includegraphics[width=0.49\textwidth]{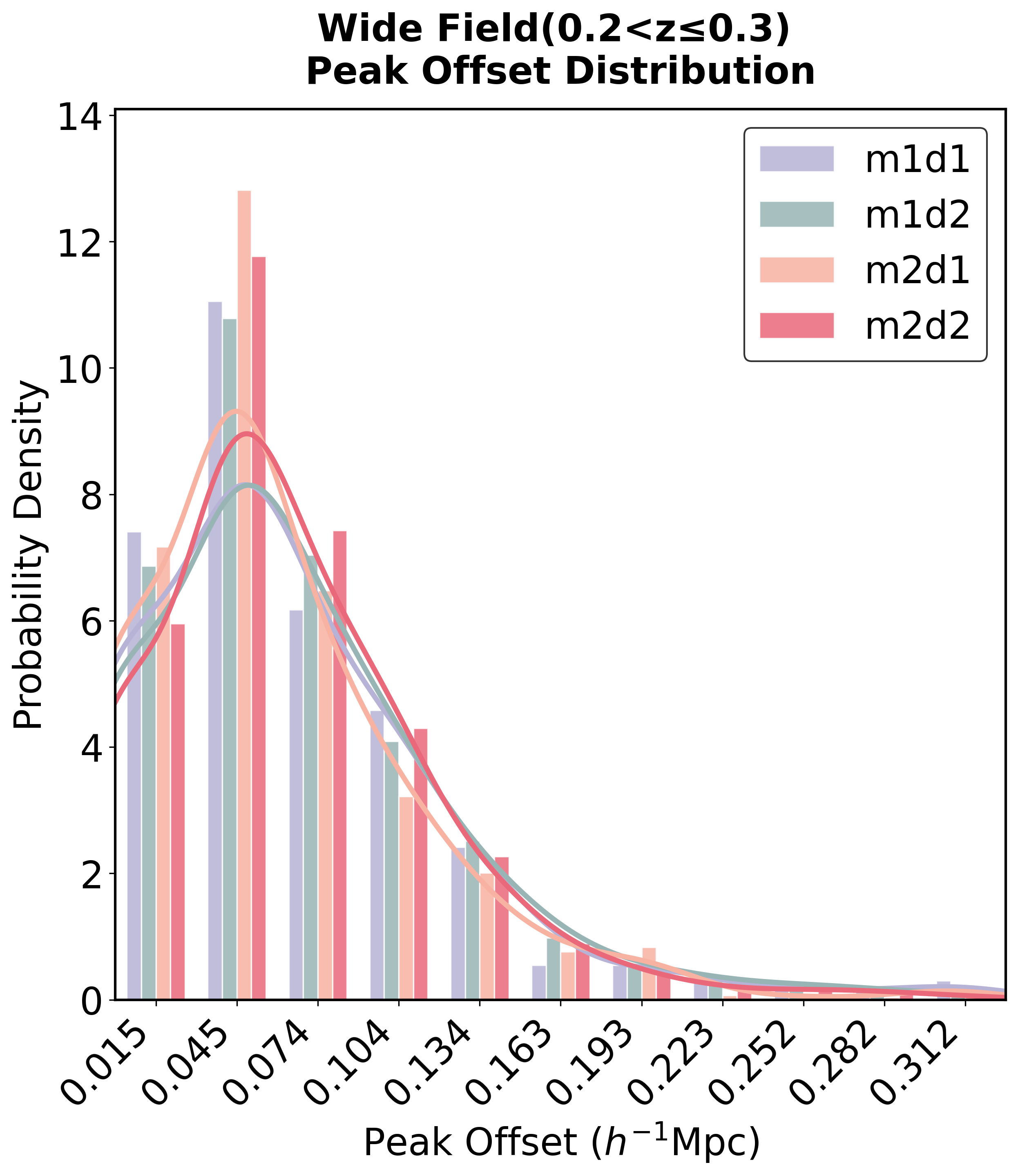}
\includegraphics[width=0.49\textwidth]{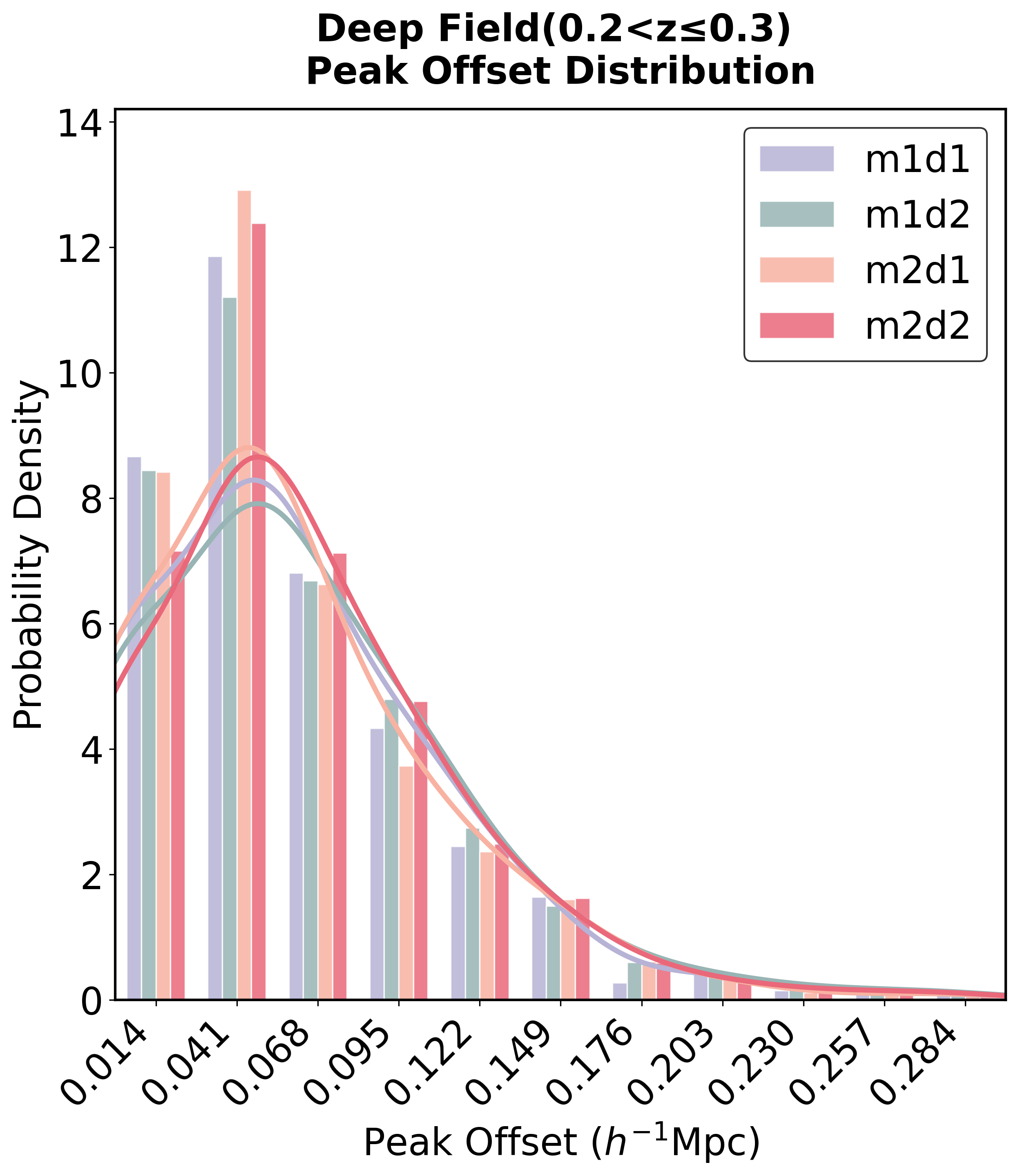}
\caption{
Distributions of projected offsets between the true and reconstructed
halo centers for the subsamples defined in Table~\ref{tab:bin}. The
$z1$ ($0.1<z\leq0.2$) and $z2$ ($0.2<z\leq0.3$) bins are presented
in the top and bottom rows, respectively, with the wide- and deep-field
cases shown separately. The histograms show the normalized
distributions in $h^{-1}\mathrm{Mpc}$, together with Gaussian KDE
curves using a bandwidth factor of 0.4. $m1$ and $m2$
denote
$14.5<\log_{10}[M/(h^{-1}M_\odot)]\leq14.8$ and
$14.8<\log_{10}[M/(h^{-1}M_\odot)]\leq15.5$, respectively, while $d1$
and $d2$ denote $d_{\rm proj}<5\,h^{-1}\mathrm{Mpc}$ and
$5\leq d_{\rm proj}<10\,h^{-1}\mathrm{Mpc}$, respectively.}
\label{fig:his} 
\end{figure*}

\subsection{Simulation}
\label{sec:Simulation}

We combine two cosmological $N$-body simulations with complementary volume and resolution. The first, corresponding to the dataset used in \citet{2026RAA....26g5003H}, provides multiple realizations of general large-scale structures for statistical robustness, while the second spans a much larger cosmological volume, improving the sampling of rare massive systems.

\subsubsection{Simulation Data 1}

We use the same dark matter N-body simulations as in \citet{2026RAA....26g5003H} to model the matter distribution, where the dataset consists of 8 pseudo-independent simulation boxes, each with a comoving side length of $320\,h^{-1}\,\mathrm{Mpc}$ and contains $640^{3}$ dark matter particles of mass $9.7 \times 10^{9}\,h^{-1}\,M_\odot$. To cover the redshift range $0 < z \leq 1$, eight snapshots at different redshifts are extracted from the same realization. Pseudo-independence among these snapshots is achieved through random translations and rotations. We generate four independent realizations in total, each with distinct initial conditions. The adopted cosmological parameters are $\sigma_{8} = 0.82$, $\Omega_{m} = 0.28$, $\Omega_{\Lambda} = 0.72$, $\Omega_{b} = 0.046$, $h = 0.7$, and $n_{s} = 0.96$.

\subsubsection{Simulation Data 2}

We additionally use a second simulation dataset derived from the \textbf{Hyper-Millennium (HM)} simulation \citep{Wang_2026}, which is a large-volume cosmological $N$-body run designed for the needs of upcoming wide-field galaxy surveys. It evolves $4.2\times10^{12}$ dark matter particles in a box with a side length of 2.5 $h^{-1}$ Gpc, providing a volume comparable to state-of-the-art simulations while maintaining high mass resolution. With 100 output snapshots, HM enables detailed tracking of structure growth and the construction of survey-ready mock catalogues, and its large volume significantly increases the number of massive halos and interacting systems, allowing us to identify a substantially larger sample of halo pairs and merging clusters that are well suited to the scientific goals of this work.

The simulation adopts a \textit{Planck} cosmology \citep{2020A&A...641A...6P} 
($H_0 = 67.66~\mathrm{km\,s^{-1}Mpc^{-1}}$, 
$\Omega_{\mathrm{m}}=0.3111$, 
$\Omega_{\Lambda}=0.6889$, 
$\Omega_{\mathrm{b}}=0.049$, 
$\sigma_8=0.8102$, 
$n_s=0.9665$).
Initial conditions are generated at $z=127$ using the Zel’dovich approximation and the linear power spectrum from \textsc{CAMB} \citep{Lewis:1999bs}  
The particle mass is $3.2\times10^{8}~h^{-1}M_\odot$, and the force softening is $\epsilon = 3~h^{-1}\mathrm{kpc}$ with a quadratic softening kernel for improved small-scale force accuracy. Halos are identified using a Friends-of-Friends algorithm with linking length $0.2$ of the mean particle separation.

In this work, the simulations adopted for training are generated using two different cosmological parameter sets. Nevertheless, we do not expect this difference to have a significant impact on the main conclusions of this study. As shown in \citet{2026RAA....26g5003H}, a model trained on simulations constructed under one cosmological framework can still achieve robust performance when applied to data generated from a different cosmological simulation. This behavior can be understood from the fact that the network primarily learns the local mapping between dark matter distributions and the corresponding weak-lensing shear signals, a relationship fundamentally determined by gravitational lensing physics and independent of the specific choice of cosmological parameters. Therefore, provided that the training data contain sufficiently diverse realizations of matter distributions and lensing configurations, the trained model is expected to retain good generalization capability and transferability across different cosmological settings.

\section{Methodology}
\label{sec:Methodology}
\subsection{Deep Learning Method}

For the reconstruction of the physical fields, we utilize a neural network based on the \textbf{Mass-Inversion-U2Net (MIU$^{2}$Net)} framework, as proposed by \citet{2026RAA....26g5003H}. This architecture builds on the nested U-Net (U$^{2}$-Net) structure of \citet{qin2020u2}, which is known for increasing network depth without significantly raising computational cost. Its large encoder-decoder structure, with residual connections at each stage, allows efficient fusion of information across multiple scales.

Following \citet{2026RAA....26g5003H}, the MIU$^{2}$Net framework incorporates several key modifications to the original U2-Net design. Specifically, the sigmoid function is removed from the final decoder stage, allowing the output to be unbounded and directly represent the convergence field. To mitigate the overly smooth predictions of the standard Mean Squared Error (MSE), we adopt a hybrid loss function,
\begin{equation}
    l = \alpha \, l_\mathrm{MSE} + \beta \, l_\mathrm{RAPS},
\end{equation}
where the MSE term enforces pixel-wise accuracy,
\begin{equation}
    l_\mathrm{MSE} = \frac{1}{N_{\rm pix}} \sum_{i,j}
    \left( \kappa_{i,j} - \hat{\kappa}_{i,j} \right)^2,
\end{equation}
with $N_{\rm pix}$ the total number of pixels, and the RAPS term constrains the statistical properties in Fourier space,
\begin{equation}
    l_\mathrm{RAPS} = \frac{1}{r_{\max}} \sum_{r=0}^{r_{\max}}
    \left| \bar{P}_{\kappa}(r) - \bar{P}_{\hat{\kappa}}(r) \right|,
\end{equation}
where $\bar{P}(r)$ denotes the radial-averaged power spectrum. In this work, we adopt $r_{\max}=16$.

Our model takes the complex reduced shear map as a two-channel input and learns a non-linear mapping to the true convergence map.

\begin{figure*}
\centering
\includegraphics[width=0.85\textwidth]{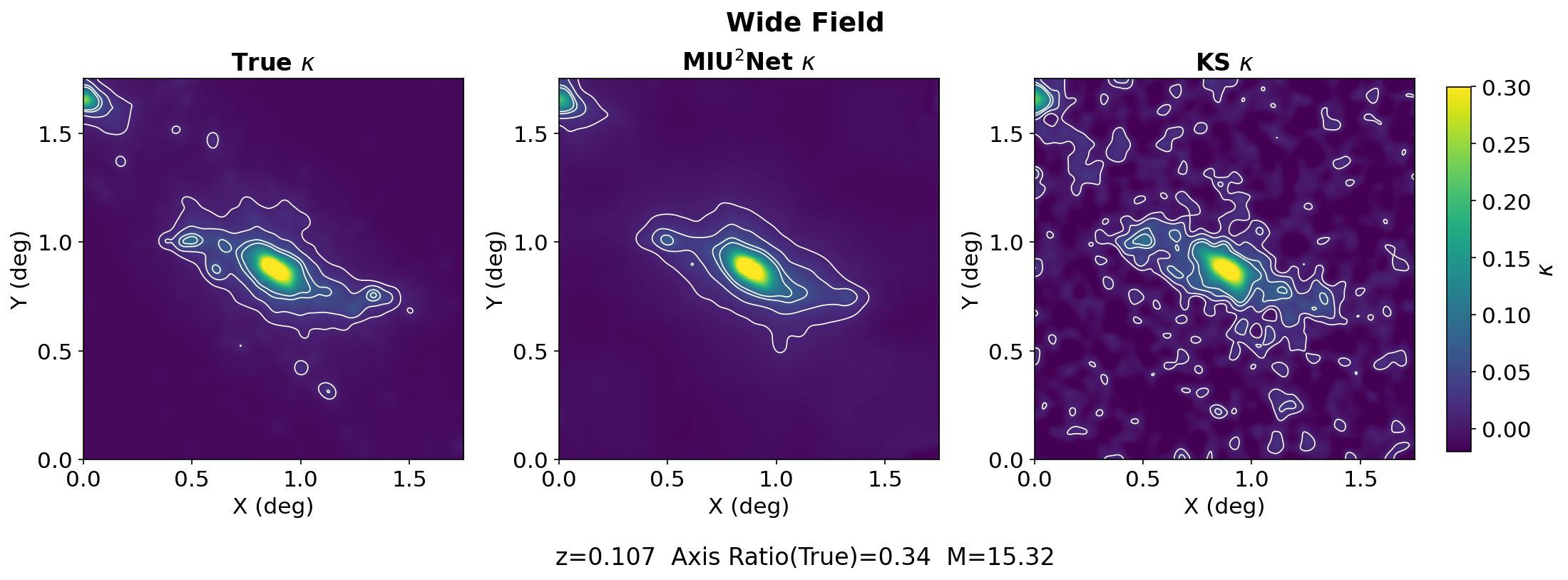}
\includegraphics[width=0.85\textwidth]{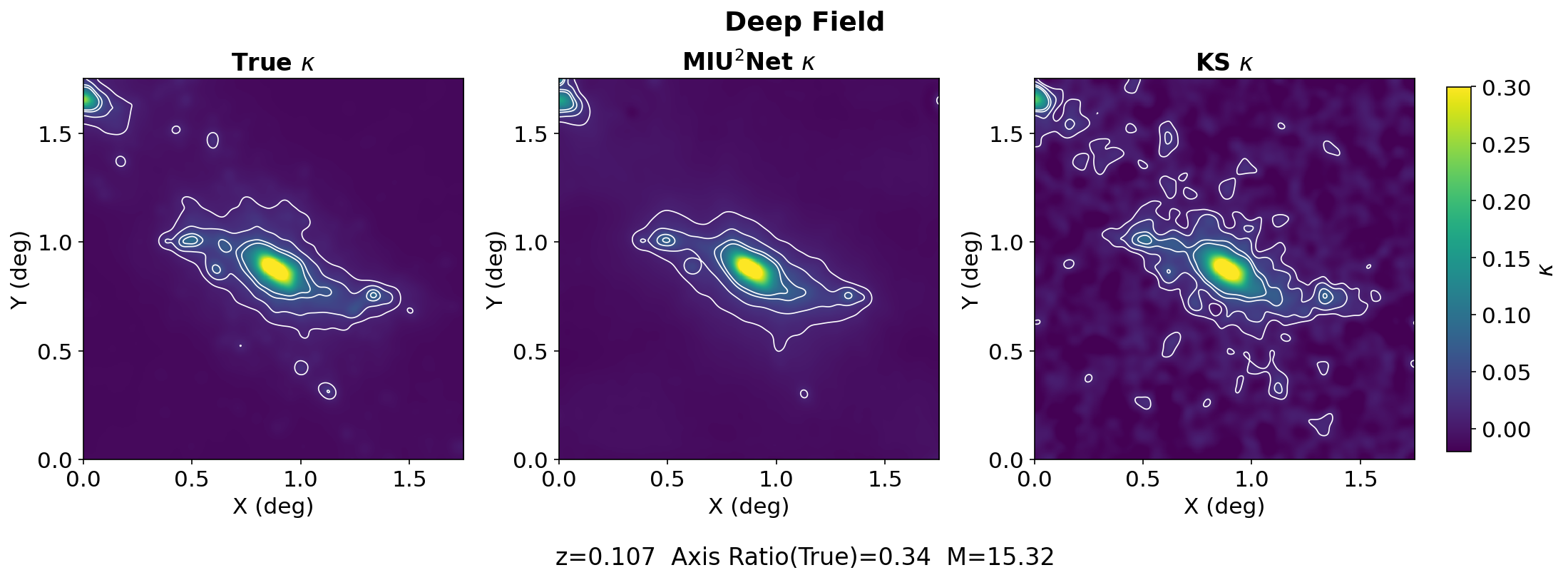}
\includegraphics[width=0.85\textwidth]{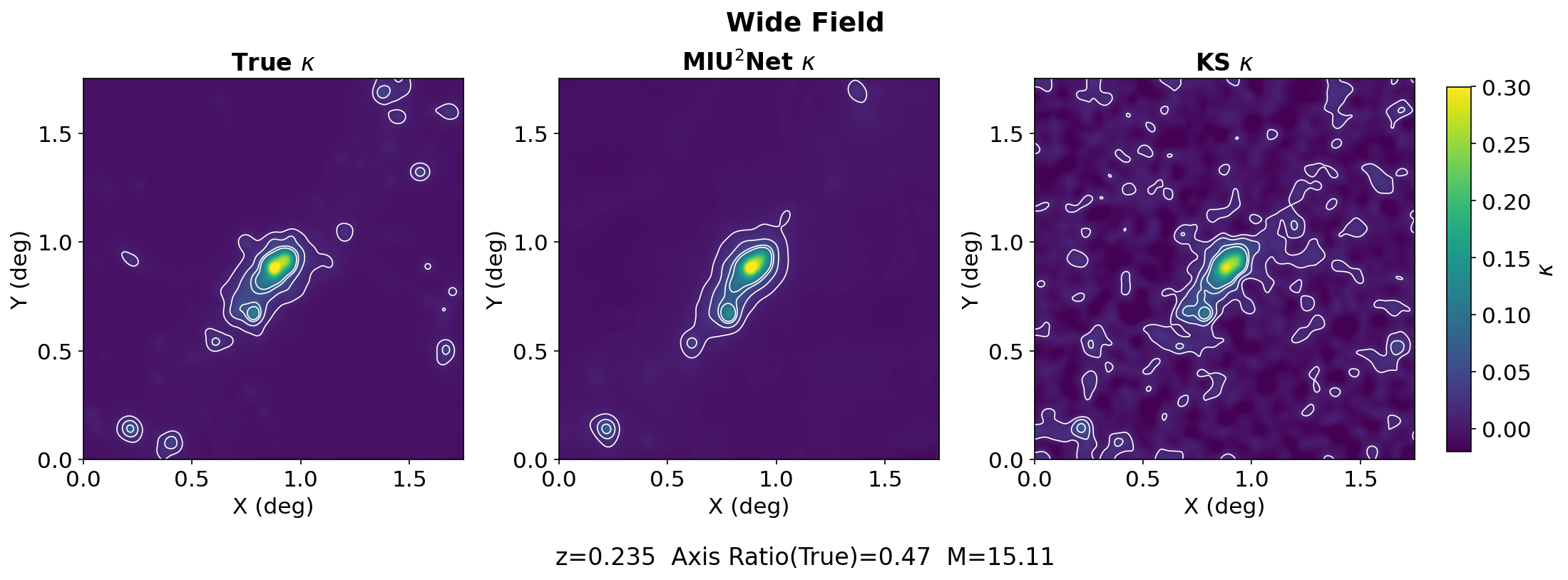}
\includegraphics[width=0.85\textwidth]{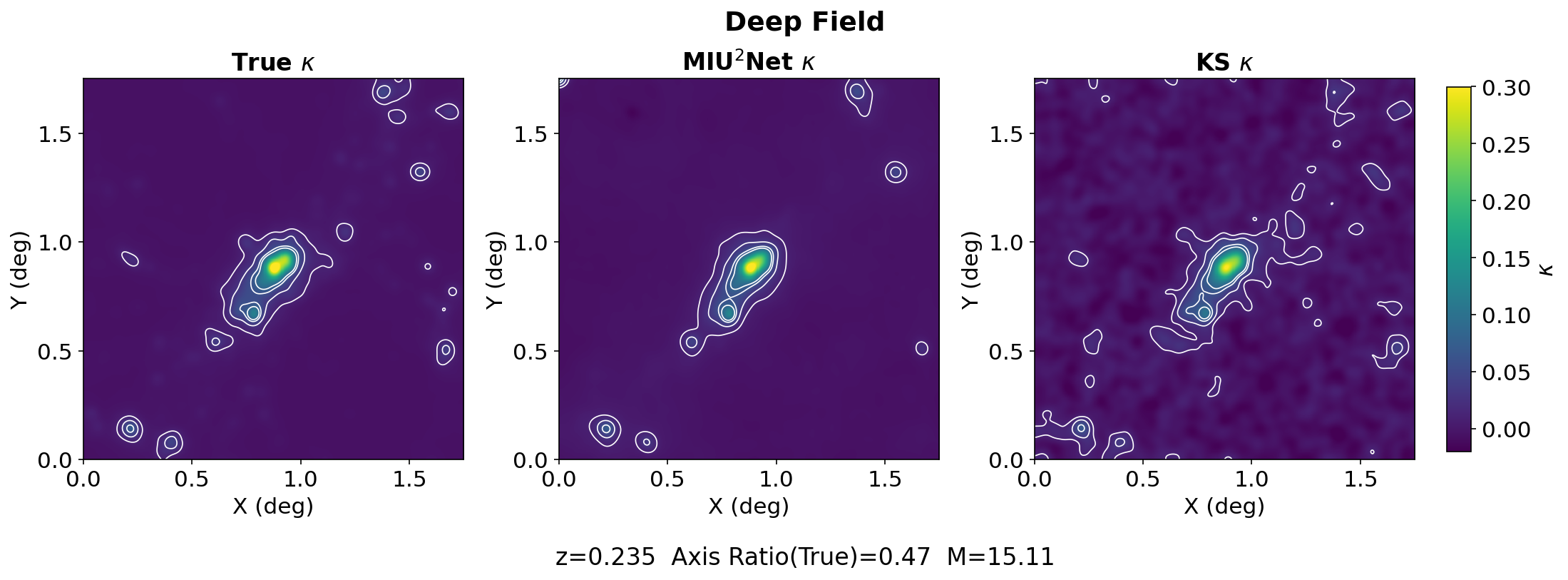}
 \caption{Two example convergence maps for single elongted-halo systems. Contours at $\kappa=[0.01, 0.03, 0.06, 0.08]$ highlight structures at different density levels. The layout is the same as in Fig.~\ref{fig:comparison1}.}
\label{fig:comparison2} 
\end{figure*}

\begin{figure*}
\centering
\includegraphics[width=0.49\textwidth]{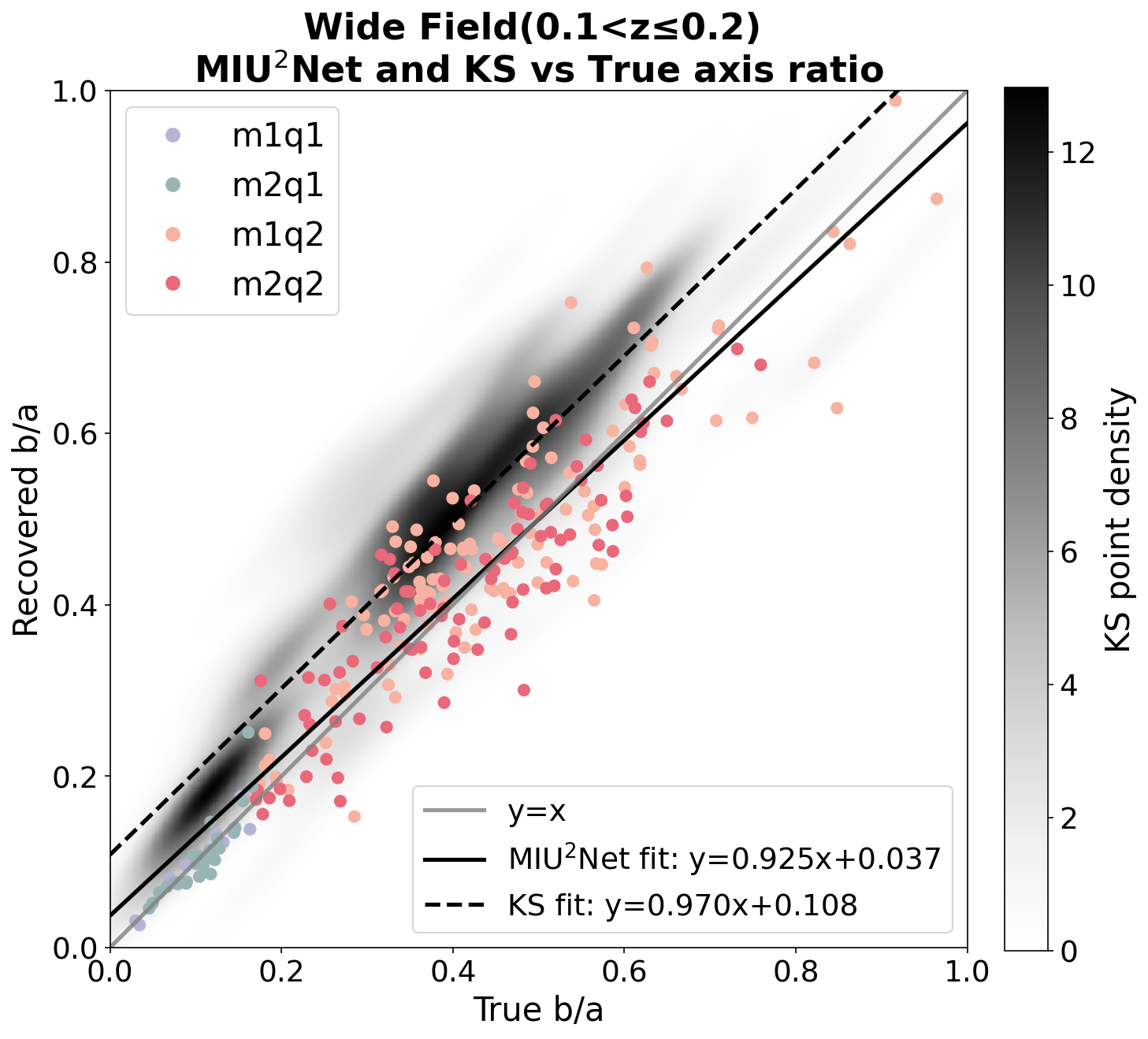}
\includegraphics[width=0.49\textwidth]{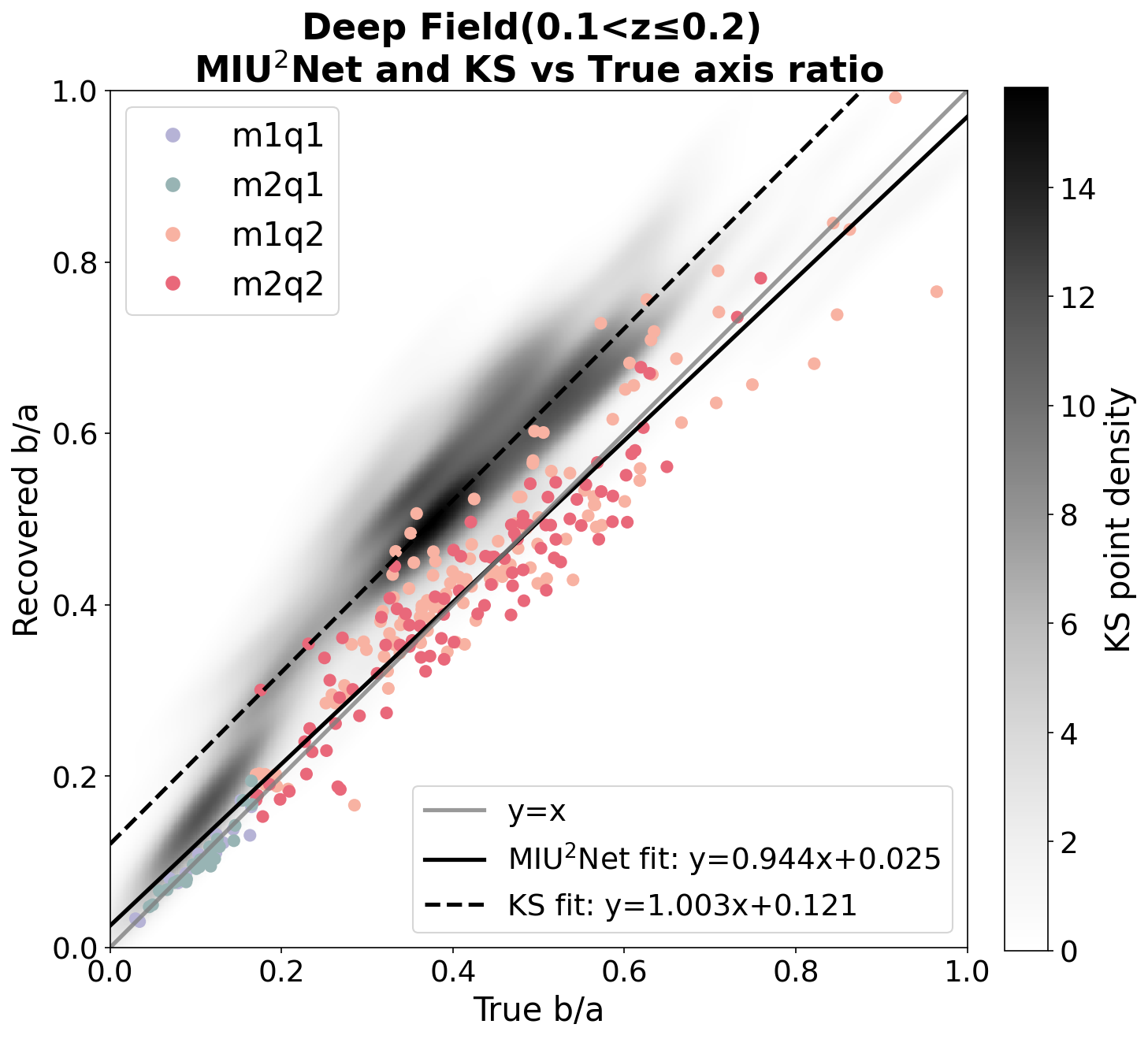}
\includegraphics[width=0.49\textwidth]{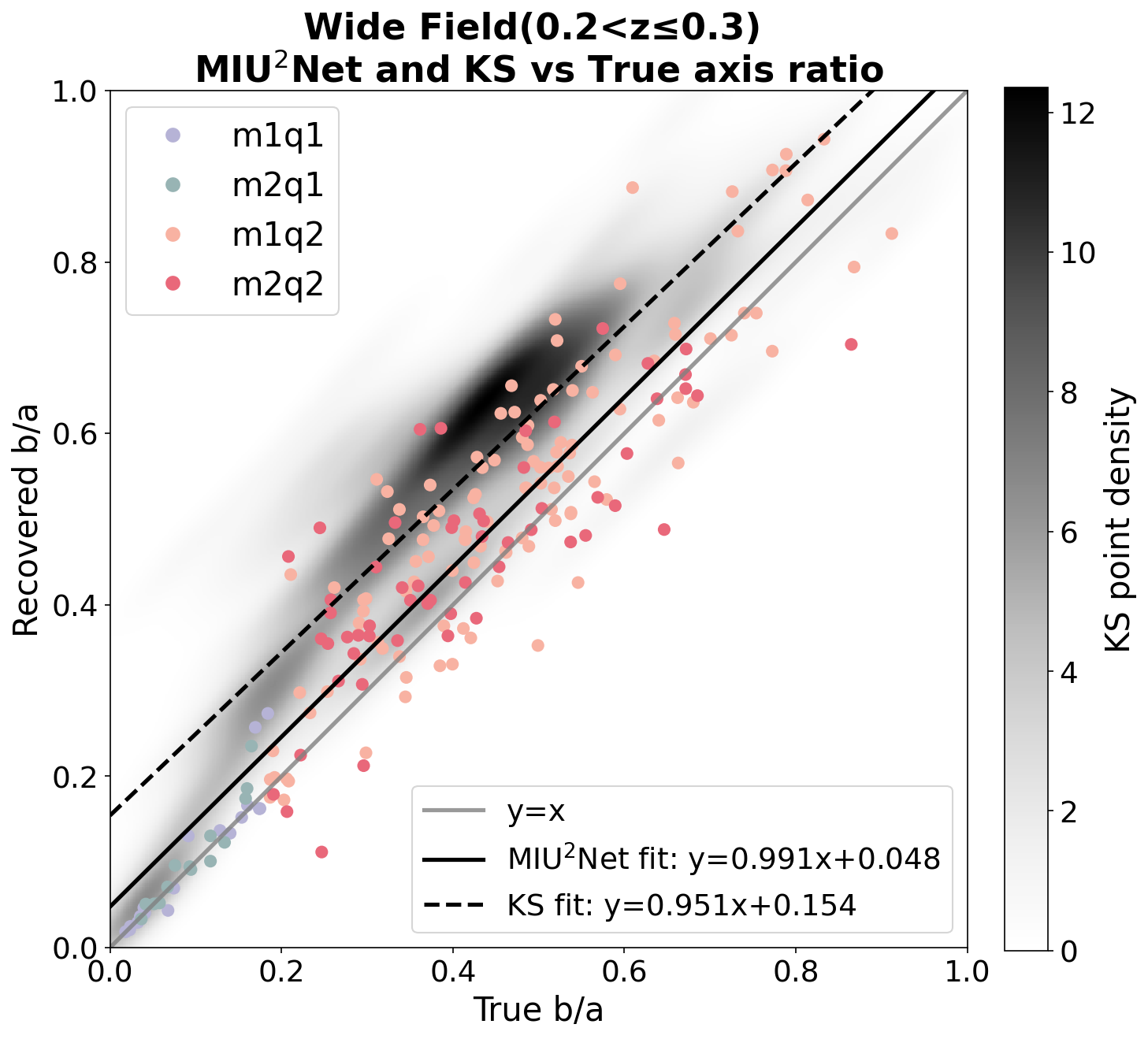}
\includegraphics[width=0.49\textwidth]{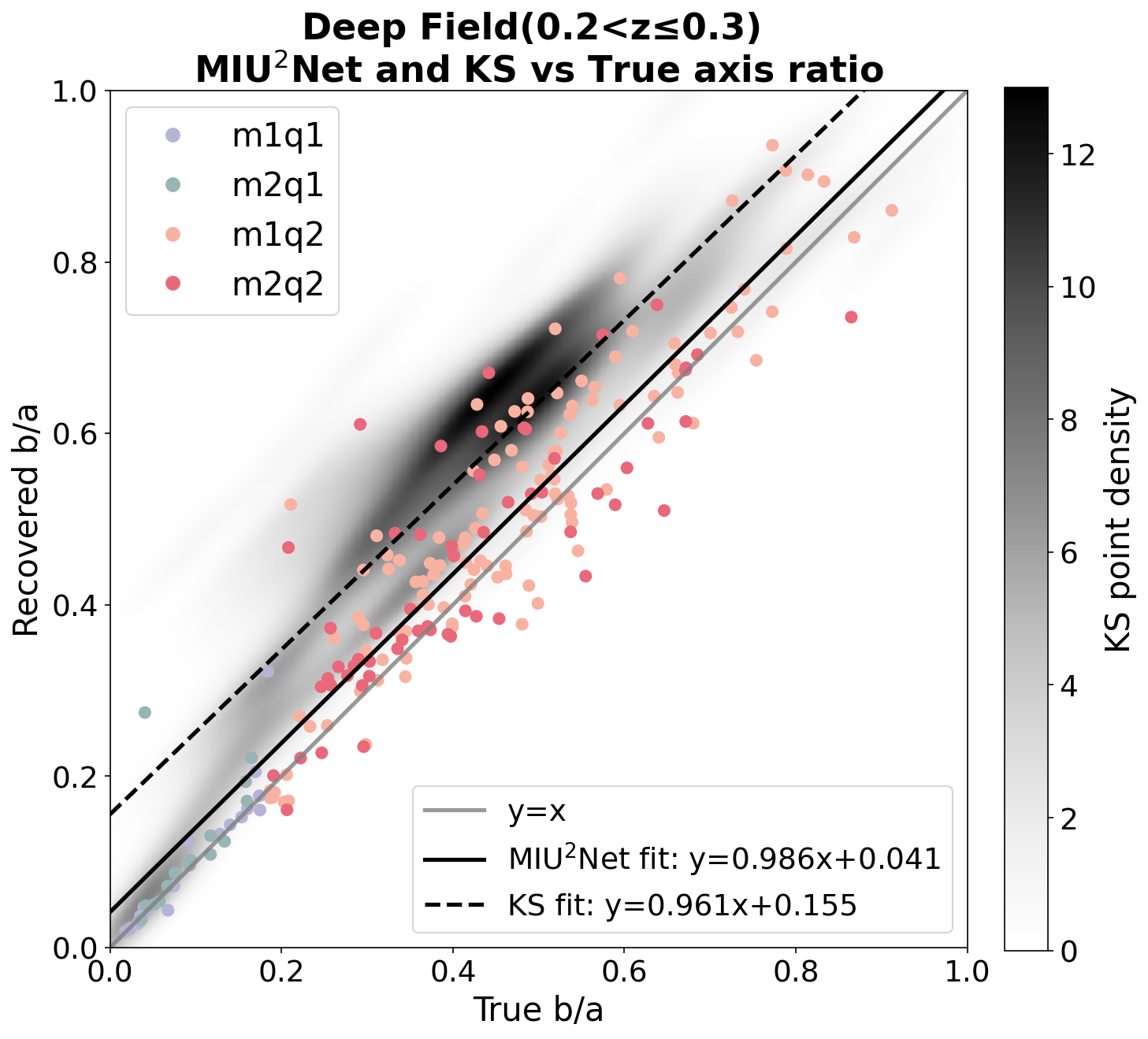}

\caption{
Comparison of the true and reconstructed halo axis ratios for the
subsamples defined in Table~\ref{tab:bin}. The $z1$
($0.1<z\leq0.2$) and $z2$ ($0.2<z\leq0.3$) bins are presented in the
top and bottom rows, respectively, with the wide- and deep-field cases
shown separately. The four colored MIU$^{2}$Net subsamples are defined
by halo mass and true projected axis ratio. Here, $m1$ and $m2$ denote
$15.0<\log_{10}[M/(h^{-1}M_\odot)]\leq15.2$ and
$15.2<\log_{10}[M/(h^{-1}M_\odot)]\leq15.5$, respectively. The $q1$
subsample contains the lowest 15\% of the true axis-ratio distribution
in each redshift bin, while $q2$ contains the remaining 85\%. Each colored point represents an individual MIU$^{2}$Net reconstruction, while the KS distribution is represented by a smoothed two-dimensional KDE with a Gaussian-kernel bandwidth factor of 0.25. The gray line denotes the
one-to-one relation, while the black solid and dashed lines show the
best-fitting relations for MIU$^{2}$Net and KS, respectively.
}

\label{fig:axis} 
\end{figure*}

\subsection{Training data}
\label{sec:Trainingdata}

To generate the training, validation and test sets, we process the two simulations described above, constructing corresponding reduced shear and convergence maps from each. 

For Simulation~1, each simulation box is divided into five segments along the line of sight, and dark matter particles in each segment are projected onto two-dimensional slices serving as lens planes. In total, 37 lens planes are obtained between $0 < z \leq 1$, and multi-plane ray-tracing \citep{2009A&A...499...31H} is applied to generate the corresponding shear ($\gamma$) and convergence ($\kappa$) maps. Each map covers $3.5 \times 3.5 \mathrm{deg}^2$ with $1024 \times 1024$ pixels. For training, we use 3000 map pairs and reserve 900 for validation, and maps are center cropped to $512 \times 512$. Data augmentation with flips and rotations expands the dataset to $\sim 3.2 \times 10^{5}$ effective pairs.

For Simulation~2, we select halo pairs with redshifts
$0.1 < z \leq 0.3$, main halo masses in the range
$10^{14.5}$--$10^{15.5}$ $h^{-1}M_\odot$, and three-dimensional
separations smaller than $10\,h^{-1}\mathrm{Mpc}$. The selected halo pairs
are further divided into subsamples according to their projected separations,
as summarized in Table~\ref{tab:bin}. The selected redshift range corresponds to systems with relatively prominent lensing signals and well-resolved large-scale structures. In principle, the same framework can be readily extended to higher-redshift systems. In addition, a set of single elongated halos with varying degrees of projected elongation is identified within
the same redshift and mass ranges. To characterize their morphology and facilitate subsequent binning by halo elongation, we quantify the projected halo shape using an iterative adaptive ellipse-fitting procedure applied to the projected $\kappa$ field. Starting from a threshold-selected region, the shape is refined by iteratively computing a weighted covariance matrix of the selected pixels and updating the best-fit ellipse accordingly. The procedure is repeated until convergence or a maximum of 20 iterations. The axis ratio is defined as $b/a$, where $a$ and $b$ are the semi-major and semi-minor axes of the final ellipse.

For both the halo-pair and elongated-halo samples, we adopt the thin-lens
approximation and treat the projected matter distribution of each target
system as a single lens plane. For all Simulation~2 samples, including the
training, validation, and test sets, the background sources are placed at
$z_s=1$. This controlled setup highlights the cluster and filament
morphologies of interest and does not include unrelated foreground or
background mass concentrations along the line of sight. For each system, we extract the corresponding noise-free convergence
($\kappa$) maps and generate shear ($\gamma$) fields following the
standard Kaiser--Squires prescription. Specifically, in Fourier space, the complex
shear is related to the convergence field through a linear kernel,
\begin{equation}
\tilde{\gamma}(\boldsymbol{k}) =
\pi^{-1}\tilde{D}(\boldsymbol{k}) \tilde{\kappa}(\boldsymbol{k}),
\qquad \boldsymbol{k} \neq 0,
\end{equation}
where $\boldsymbol{k}$ denotes the Fourier wavevector. The kernel
$\tilde{D}(\boldsymbol{k})$ is defined as
\begin{equation}
\tilde{D}(\boldsymbol{k}) = \pi
\frac{k_1^2 - k_2^2 + 2 i k_1 k_2}{|\boldsymbol{k}|^2},
\end{equation}
which allows the convergence field to be recovered through the inverse
relation
\begin{equation}
\tilde{\kappa}(\boldsymbol{k}) =
\pi^{-1}\tilde{\gamma}(\boldsymbol{k})\tilde{D}^*(\boldsymbol{k})
\qquad \boldsymbol{k} \neq 0 .
\end{equation}

Finally, the shear field is converted to the reduced shear field according to Eq.~(\ref{eq:reduced}). In total, 2000 maps are used for training and 600 maps for validation for Simulation~2. Map cropping and data augmentation are implemented in the same manner as described for Simulation~1.

Overall, the final dataset combines samples from both Simulation~1 and Simulation~2. The training set consists of 3000 maps from Simulation~1, 1000 maps of halo pairs from Simulation~2, and 1000 maps of elongated merging halos from Simulation~2, yielding a total of 5000 training maps. The validation set contains 900, 300, and 300 maps from the corresponding subsets, respectively, for a total of 1500 validation maps. All maps are noise-free and serve as the basis for the shape-noise realizations described below.

\subsection{Shape-noise Realizations}

In this section, we generate mock noisy reduced shear maps to assess our reconstruction method under realistic observational conditions. We consider two source-density scenarios. For the fiducial case, we adopt \(n_g=20~\mathrm{arcmin}^{-2}\), consistent with the expected source density of the CSST wide-field weak-lensing survey \citep{liu2023potential}. We also investigate a deeper observational scenario motivated by the planned CSST deep-field survey, which will cover \(\sim 400~\mathrm{deg}^{2}\) and reach at least one magnitude deeper than the wide-field survey through repeated observations \citep{2026SCPMA..6939501C}. We estimate the corresponding source density using the COSMOS2015 photometric catalogue, following the setup of \citet{liu2023potential}. The COSMOS \(i^+\)-band magnitude is used as a proxy for the CSST \(i\) band. The CSST deep-field \(5\sigma\) limiting magnitude is converted to an approximate \(S/N>10\) threshold, and a weak-lensing source sample is then constructed by requiring this magnitude threshold and photometric redshifts in the range \(0.2<z_{\rm phot}<3.0\). This gives
\begin{equation}
n_g^{\rm CSST-deep}\simeq 47.3~\mathrm{arcmin}^{-2},
\end{equation}
which we adopt as a representative deep-field case alongside the fiducial wide-field value.

Gaussian shape noise is then added independently to each pixel. Following \citet{2004MNRAS.350..893H} and adopting the per-component convention used in COSMOS weak-lensing analyses, the noise variance for each shear component is written as
\begin{equation}
\sigma_n^2 =
\frac{\sigma_{\tilde{\gamma}}^2}{\theta_s^2 n_g},
\end{equation}
where \(\theta_s=0.21~\mathrm{arcmin}\) is the pixel size and \(n_g\) is the source galaxy number density per square arcmin. The total per-component shape noise, \(\sigma_{\tilde{\gamma}}\), includes both the intrinsic shape noise and the shape measurement error,
\begin{equation}
\sigma_{\tilde{\gamma}}^2
=
\sigma_{\rm int}^2+\sigma_{\rm meas}^2 .
\end{equation}

Following COSMOS weak-lensing analyses \citep{2007ApJS..172..219L}, we adopt \(\sigma_{\rm int}=0.26\) per component. For our \(S/N>10\) source selection, corresponding approximately to \(\mathrm{MAG\_AUTO}\simeq25.5\), we conservatively adopt \(\sigma_{\rm meas}=0.18\) per component, based on the measurement errors reported for galaxies of this magnitude in their Fig. 17. This is a conservative estimate, as galaxies with higher \(S/N\) are expected to have smaller measurement errors.

We generate noisy realizations corresponding to both the wide-field ($n_g=20~\mathrm{arcmin}^{-2}$) and deep-field ($n_g=47.3~\mathrm{arcmin}^{-2}$) galaxy number densities. Independent models are trained and evaluated for each observational scenario.

\begin{figure*}
\centering
\includegraphics[width=0.49\textwidth]{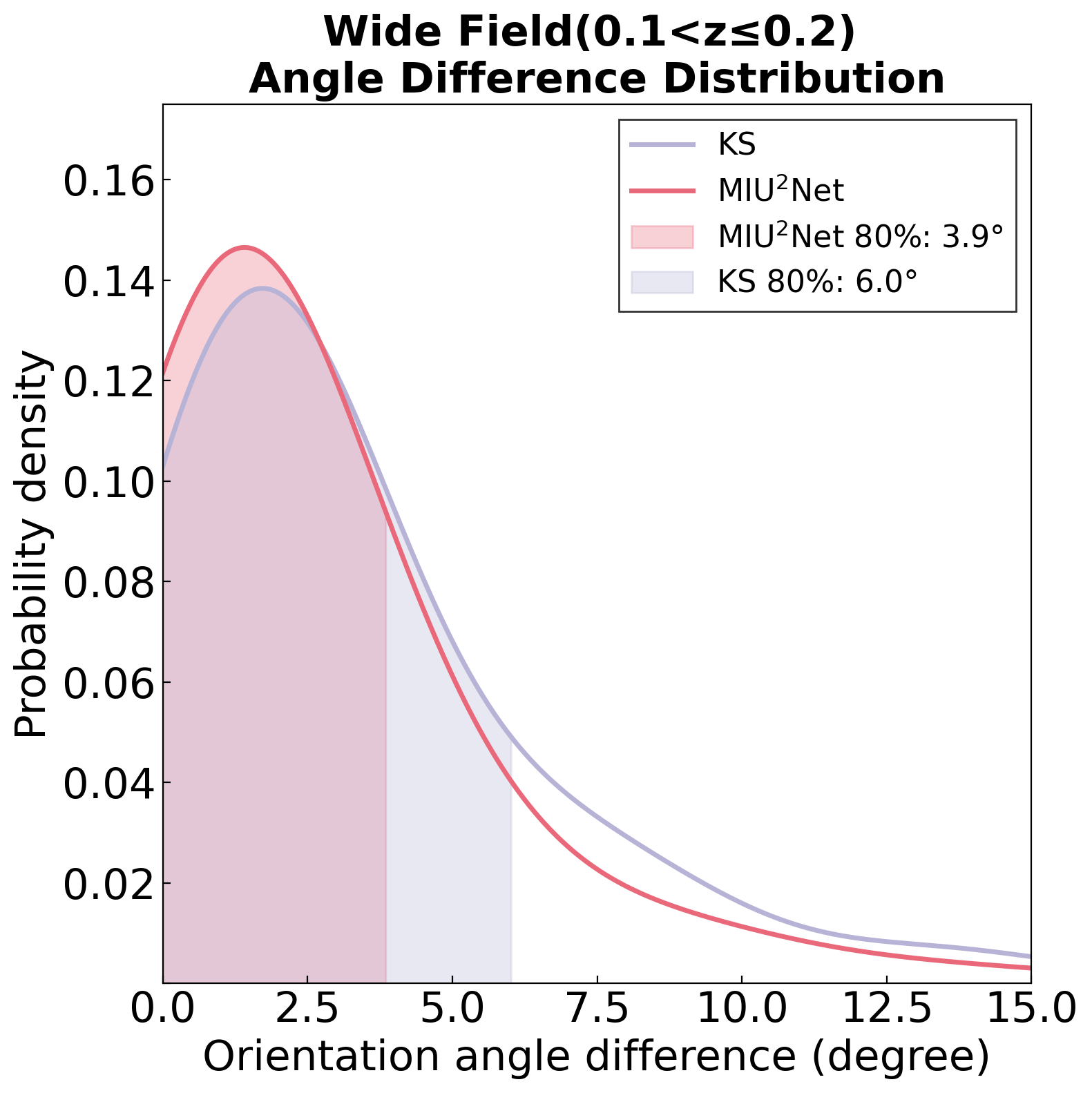}
\includegraphics[width=0.49\textwidth]{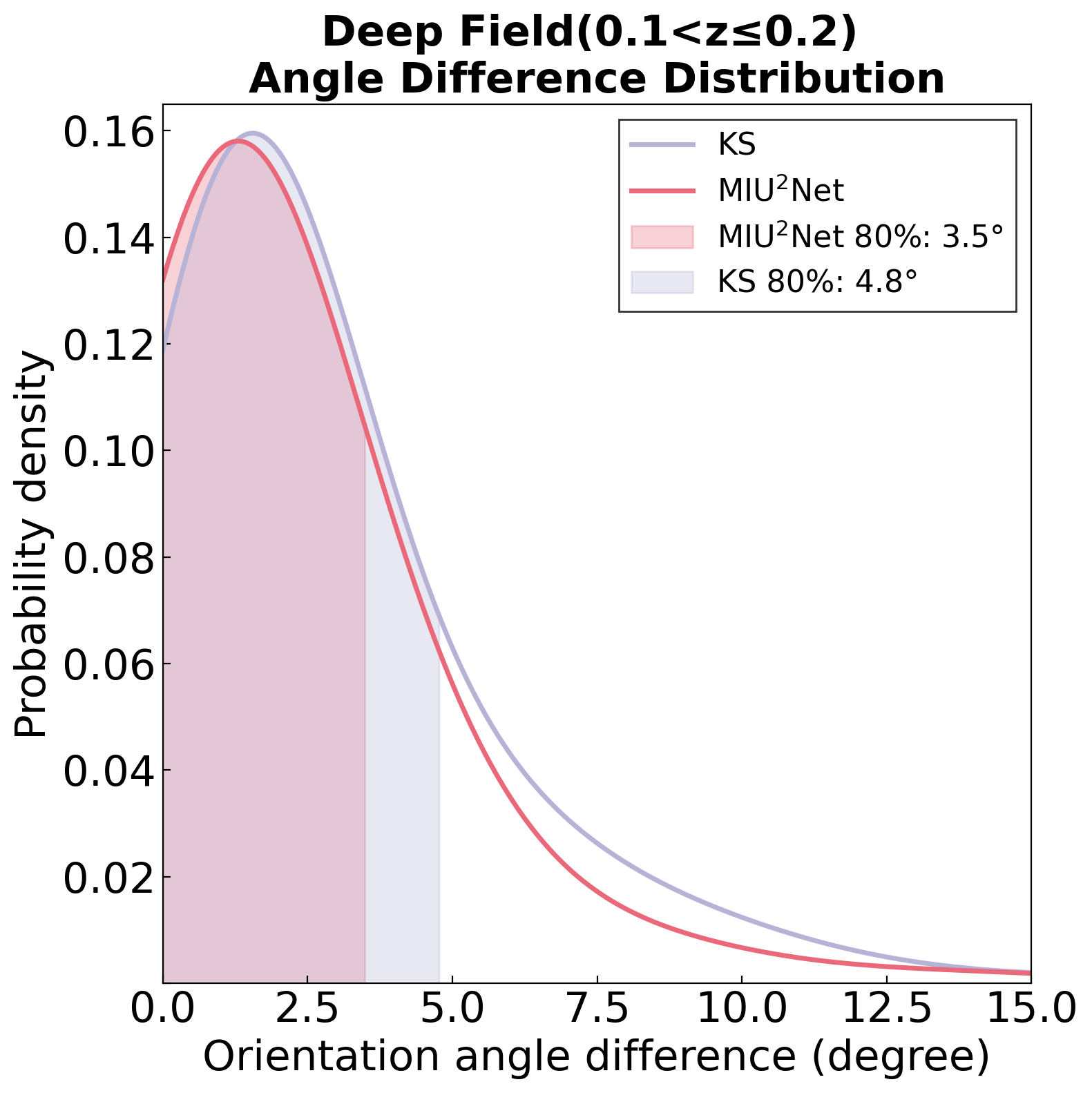}
\includegraphics[width=0.49\textwidth]{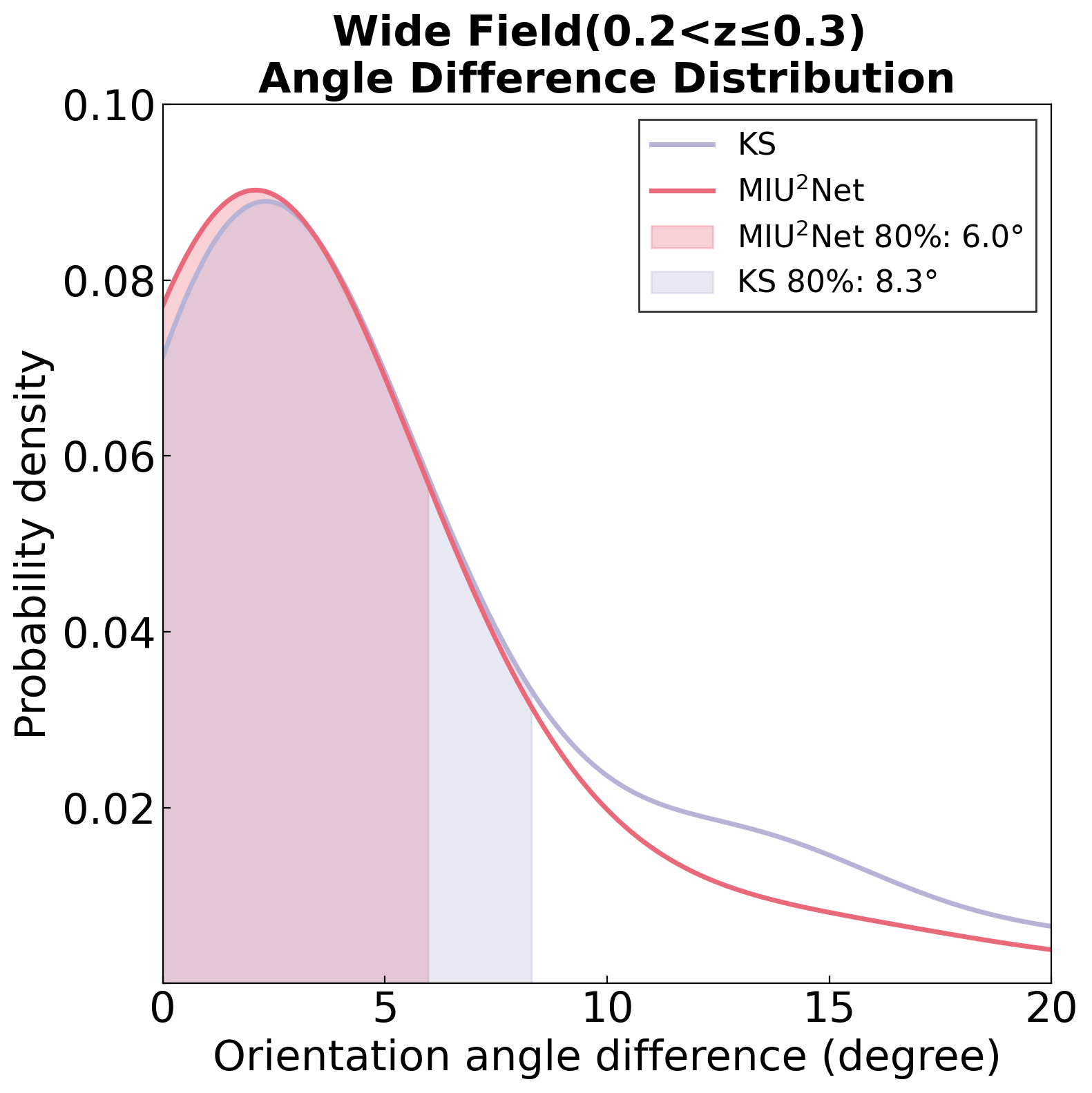}
\includegraphics[width=0.49\textwidth]{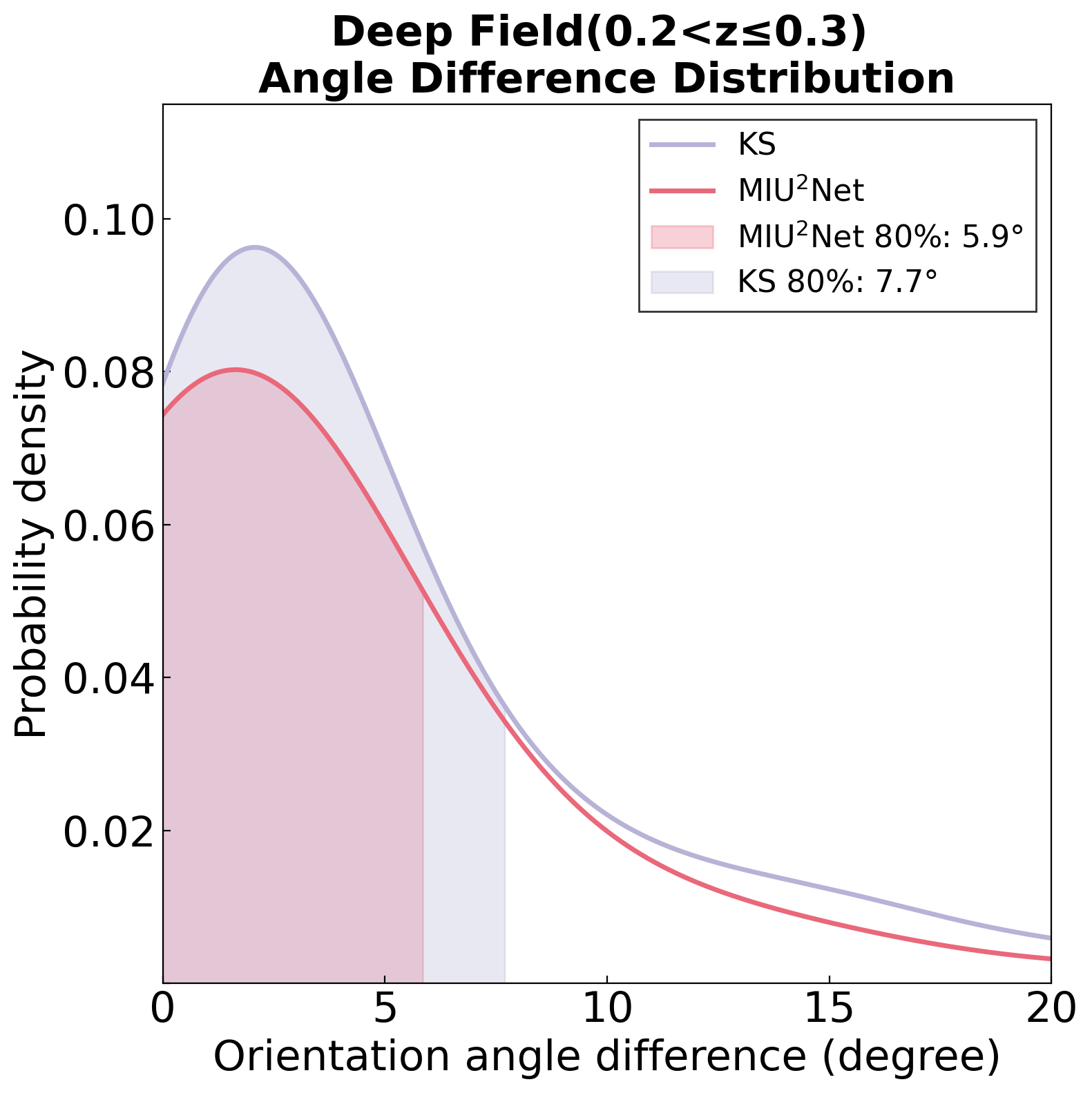}
\caption{ Distributions of the differences between the semi-major-axis
orientations measured from the reconstructed and true convergence maps
for the subsamples defined in Table~\ref{tab:bin}. The $z1$
($0.1<z\leq0.2$) and $z2$ ($0.2<z\leq0.3$) bins are presented in the
top and bottom rows, respectively, with the wide- and deep-field cases
shown separately. The orientation differences are measured in degrees
and represented by KDE curves with a bandwidth factor of 0.3.
MIU$^{2}$Net and KS are shown in red and purple, respectively. The
shaded regions indicate orientation differences below the 80th
percentile of each distribution.
}

\label{fig:angle} 
\end{figure*}

\section{Results}
\label{sec:result}

We emphasize that our analysis is restricted to two morphological
classes that are particularly intriguing for weak-lensing
reconstruction, with the corresponding test samples constructed exclusively from Simulation~2: halo pairs and elongated single halos. The halo-pair and elongated-halo samples used in the following analyses are fully independent of the training and validation sets, ensuring an unbiased evaluation of the reconstruction performance. Owing to their distinct physical configurations, these two classes are subdivided into bins using different criteria. For halo pairs, the test samples are binned according to redshift, main halo mass, and projected separation between the two halo centers. For elongated halos, we instead bin the samples by
redshift, halo mass, and halo shape, using the projected axis ratios
($b/a$) measured with the ellipse-fitting procedure described in
Sec. \ref{sec:Trainingdata}. In each redshift bin, the lowest 15\% of the true axis-ratio
distribution is classified as highly elongated ($q1$), while the
remaining 85\% forms the comparison sample ($q2$). A summary of all subsample definitions and their basic statistics is
presented in Table~\ref{tab:bin}. We note that the training and validation datasets also span the full range of bins considered in this work, with comparable numbers of samples in each bin to avoid introducing biases related to sample imbalance.

\subsection{Reconstruction of Halo Pairs} 
\label{sec:Contour}

In Fig.~\ref{fig:comparison1} we provide a side-by-side inspection of reconstructed convergence maps with contours. The comparison includes the simulated ground-truth maps, the MIU$^{2}$Net reconstructions for both wide- and deep-field cases, and those obtained with the KS approach performed using the shear field, representing the idealized limit of a converged iterative reduced-shear correction.

Overall, the MIU$^{2}$Net reconstructions show strong morphological consistency with the simulated convergence fields. Extended filamentary structures are recovered with good spatial continuity, and compact high-density peaks are well reproduced in both shape and amplitude. This behavior is seen in both the wide- and deep-field cases, while the contour shapes in the deep-field maps appear more consistent with those in the true maps, owing to the lower noise level. The KS reconstructions, as a traditional baseline, capture the large-scale features of the field and provide a useful reference for overall structure, while MIU$^{2}$Net offers improved detail preservation at smaller scales.

We compare the halo-center locations of both component halos in the simulated and MIU$^{2}$Net-reconstructed convergence maps, finding that the resulting offset distribution is strongly concentrated at small separations (Fig.~\ref{fig:his}). The resulting dispersion remains small for both wide- and deep-field cases, with the $1\sigma$ range confined within 0.08 $\,h^{-1}\,\mathrm{Mpc}$ for z1 bin ($0.1<z\leq0.2$) and 0.1 $\,h^{-1}\,\mathrm{Mpc}$ for z2 bin ($0.2<z\leq0.3$), with only a minor tail extending toward larger separations. This behavior shows little dependence on halo mass, projected separation, or redshift. These results indicate that, across different observing depths, the reconstructed $\kappa$ maps not only recover the large-scale mass distribution, but also localize individual halo centers with high precision, providing a robust tracer of cluster locations.

\subsection{Reconstruction of Elongated Halos}
\label{sec:Quality}

We evaluate the reconstruction of elongated halo morphology using both visual inspection and quantitative shape measurements. A direct visual inspection of the reconstructed convergence maps for elongated halos, along with their contours, is provided in Figure~\ref{fig:comparison2}. For halos grouped by their true axis ratios, we measure the corresponding ratios from the reconstructed maps using both MIU$^{2}$Net and KS methods, following the same procedure described in Section~\ref{sec:Trainingdata}, and compare them against the true values in Figure~\ref{fig:axis}. The linear fits are performed with $3\sigma$ clipping to mitigate the impact of outliers that arise primarily from instabilities in the ellipticity estimation procedure (e.g., poorly defined shapes in noisy or irregular reconstructions). The slope values remain close to unity for both methods in both redshift bins, with values of 0.925/0.944 (MIU$^{2}$Net) and 0.970/1.003 (KS) in the z1 bin ($0.1<z\leq0.2$) for the wide-/deep-field cases, and 0.991/0.986 (MIU$^{2}$Net) and 0.951/0.961 (KS) at z2 bin ($0.2<z\leq0.3$), respectively. The deep-field results follow the same overall axis-ratio relation as the wide-field case. A clearer distinction emerges when considering the average absolute bias between the reconstructed and true axis ratios. 
MIU$^{2}$Net achieves consistently smaller average absolute biases,
with values of 0.049/0.038 (MIU$^{2}$Net) and
0.108/0.124 (KS) in the z1 bin ($0.1<z\leq0.2$) for the
wide-/deep-field cases, and 0.068/0.057 (MIU$^{2}$Net) and
0.139/0.141 (KS) in the z2 bin ($0.2<z\leq0.3$), respectively.
 These results indicate that MIU$^{2}$Net more closely reproduces the true halo morphology while maintaining the overall trends recovered by both methods. We further compute the Pearson correlation coefficient between the reconstructed and true axis ratios,
\begin{equation}
R =
\frac{\sum_i (q^{\rm rec}_i-\bar{q}^{\rm rec})(q^{\rm true}_i-\bar{q}^{\rm true})}
{\sqrt{\sum_i (q^{\rm rec}_i-\bar{q}^{\rm rec})^2}
 \sqrt{\sum_i (q^{\rm true}_i-\bar{q}^{\rm true})^2}},
\end{equation}
where $q=b/a$. At z1 bin, the $R$ values are 0.936 (wide field) and 0.960 (deep field) for
MIU$^{2}$Net, compared with 0.907 and 0.932 for KS,
respectively; at z2 bin, the corresponding values are 0.927 and 0.936
for MIU$^{2}$Net, compared with 0.900 and 0.914 for KS.

We further quantify halo orientation recovery using the same best-fitting ellipses employed for the axis-ratio measurements described in Section~3.2. The position angle is defined as the direction of the semi-major axis of the final fitted ellipse. Taking the orientations measured from the true convergence maps as the reference, we calculate the position-angle offsets of the reconstructed elongated halos. Across all four KDE distributions, MIU$^{2}$Net is systematically more concentrated toward small angle differences than KS, indicating a more stable recovery of the fitted major-axis orientation. For the z1/z2 bins, the 80th-percentile offsets of MIU$^{2}$Net are $3.9^\circ/6.0^\circ$ in the wide-field case and $3.5^\circ/5.9^\circ$ in the deep-field case, while the corresponding KS values are $6.0^\circ/8.3^\circ$ and $4.8^\circ/7.7^\circ$, respectively. MIU$^{2}$Net therefore maintains a narrow distribution in both
observing-depth cases, with only a mild improvement in the deep field.
The remaining large-offset tail is mainly associated either with
nearly circular fitted ellipses ($b/a\simeq1$), for which the
major-axis angle is intrinsically poorly constrained, or with
limitations of the ellipse-fitting procedure in cases with ambiguous
morphologies.

Overall, KS introduces a stronger additive bias in axis ratio and larger deviations in halo orientation, likely due to noise amplification and the lack of regularization in the direct Fourier inversion, which effectively suppresses small-scale anisotropic features and leads to overly smoothed reconstructions, whereas MIU$^{2}$Net suffers from a milder version likely arising from the network's tendency to regress toward mean morphologies. These trends are qualitatively unchanged when moving from the wide- to the deep-field setting, suggesting that deeper observations mainly improve the stability of the reconstructed halo shape rather than fundamentally altering the morphology bias.

\begin{figure*}
\centering
\includegraphics[width=0.9\textwidth]{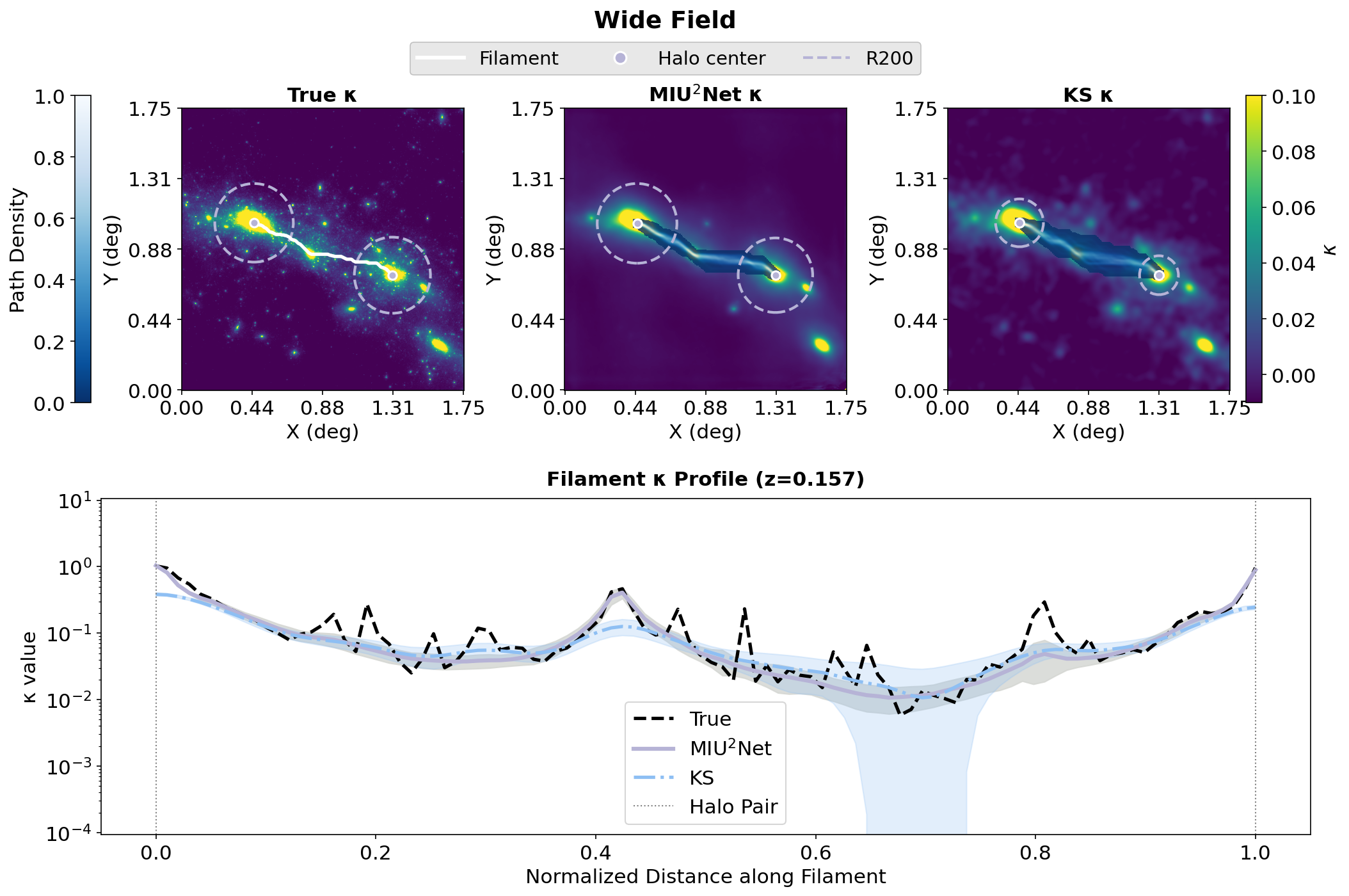}
\includegraphics[width=0.9\textwidth]{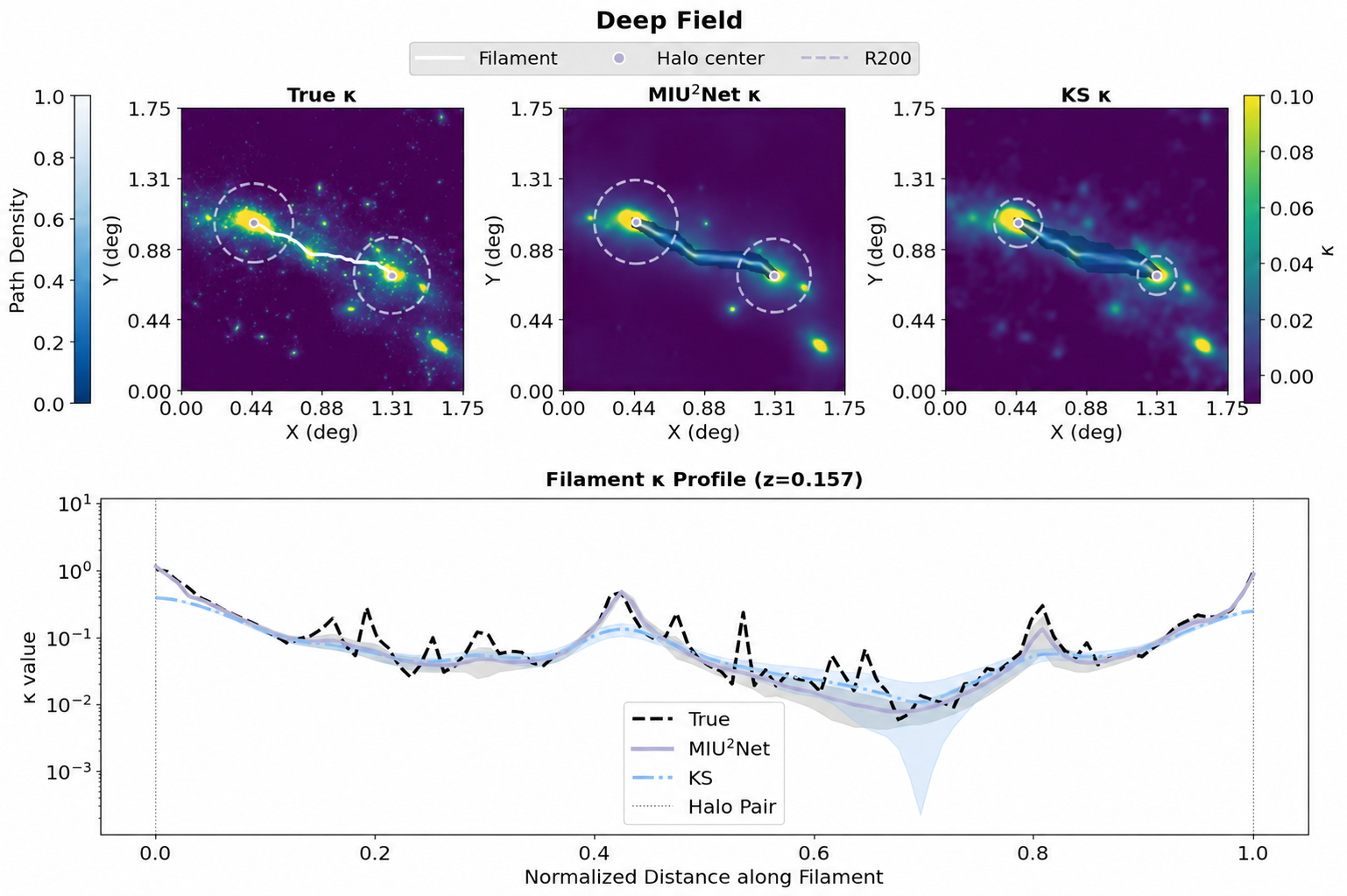}
\caption{An example of filament-finder result in the z1 bin ($0.1<z\leq0.2$) for the corresponding wide- and deep-field cases. Top row: filament identified on the true $\kappa$ map (left), and the normalized path-probability maps derived from 50 noise realizations and overlaid on the corresponding mean $\kappa$ maps for MIU$^{2}$Net (middle) and KS (right). All three maps have a pixel size of $0.21$ arcmin. Bottom row: convergence profiles between the two halos. The black dashed line shows the true profile, while the purple solid and blue dashed lines show the mean profiles derived from 50 MIU$^{2}$Net and KS realizations, respectively, with the corresponding $1\sigma$ dispersions shown as shaded regions. The distance is normalized by the total length of each filament path.}

\label{fig:filament_z0102} 
\end{figure*}

\begin{figure*}
\centering
\includegraphics[width=0.9\textwidth]{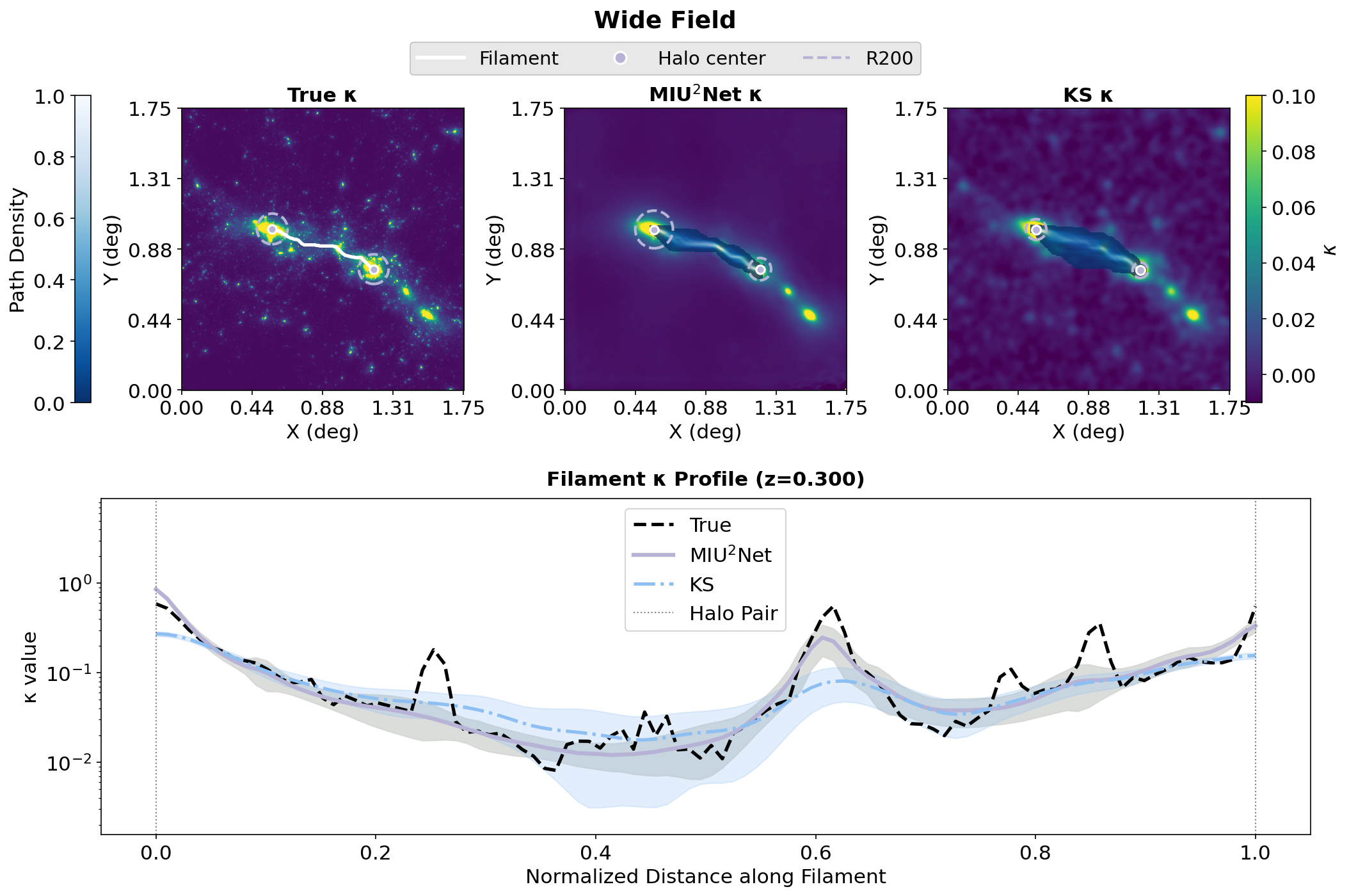}
\includegraphics[width=0.9\textwidth]{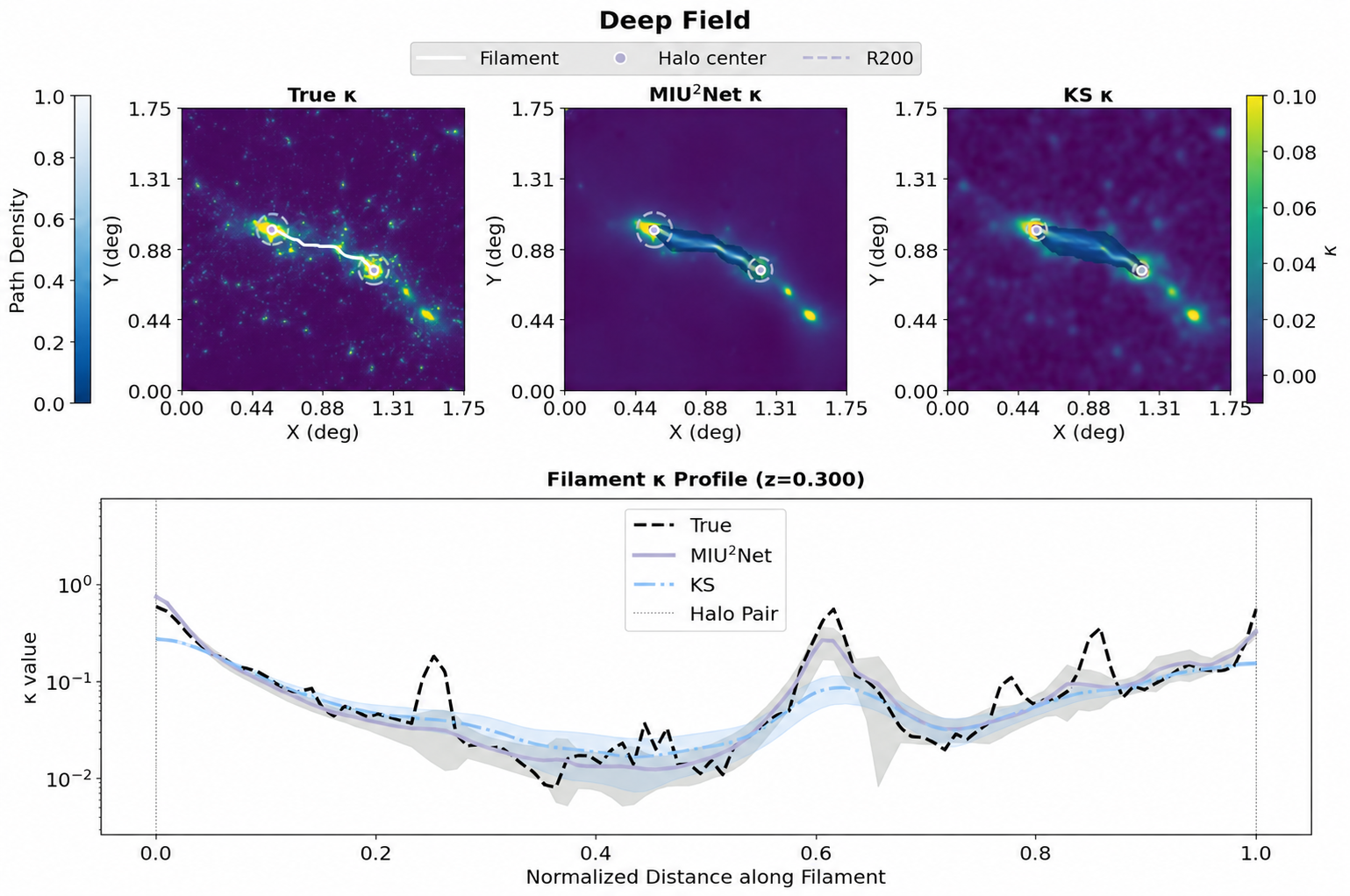}
\caption{An example of filament-finder result in the z2 bin ($0.2<z\leq0.3$) for the corresponding wide- and deep-field cases. The layout is the same as in Fig.~\ref{fig:filament_z0102}.}

\label{fig:filament_z0203} 
\end{figure*}

\subsection{Filament Path}
\label{subsec:filament-finder}

We develop an automated pipeline to identify and characterize filamentary structures connecting dark matter halos in weak-lensing convergence ($\kappa$) maps. The method integrates halo peak identification, cost-based path optimization, bootstrap-based robustness assessment, and physical property estimation within a unified framework. We apply this pipeline to a set of halo pairs with physical separations spanning 0-10 $h^{-1}$ Mpc, as summarized in Table~1.

\paragraph{Filament Path Optimization}
For each halo pair $(H_a,H_b)$, the filament is identified as a
continuous path connecting the two halos along regions of convergence.
A ``cost map'' is constructed by taking the negative of the convergence
field, $c(\boldsymbol{x})=-\kappa(\boldsymbol{x})$, so that higher
convergence corresponds to lower cost and paths are encouraged to
follow dense regions. The cost map is then linearly rescaled using
min--max normalization,
\begin{equation}
C(\boldsymbol{x})
=
\frac{c(\boldsymbol{x})-c_{\min}}{c_{\max}-c_{\min}}
=
\frac{\kappa_{\max}-\kappa(\boldsymbol{x})}
{\kappa_{\max}-\kappa_{\min}}.
\end{equation}
This normalization produces a non-negative cost map and ensures
numerical stability in the shortest-path computation. The filament
spine is obtained by tracing the minimum-cost path connecting the two
halos on an eight-connected pixel grid, which naturally follows the
high-convergence ridge between them.

\paragraph{Bootstrap Robustness Analysis}
To assess robustness against noise fluctuations, we perform a bootstrap analysis using $N=50$ independent noise realizations. For each realization, an optimal filament path is computed, yielding a set of predicted paths $\{\mathcal{P}_i\}$. Their spatial consistency is quantified through a path probability map,
\begin{equation}
P(x,y) = \frac{1}{N}\sum_{i=1}^{N} \mathbb{I}\big[(x,y)\in\mathcal{P}_i\big],
\end{equation}
where $\mathbb{I}[\cdot]$ denotes the indicator function, defined as $\mathbb{I}[A]=1$ if the condition $A$ is satisfied and $\mathbb{I}[A]=0$ otherwise. This map represents the fraction of realizations in which the filament passes through each pixel. Path density maps (Fig.~\ref{fig:filament_z0102}, upper-middle and upper-right panels) is then derived to characterize the spatial dispersion and uncertainty of the detected filament.

\begin{figure*}
\centering
\includegraphics[width=0.49\textwidth]{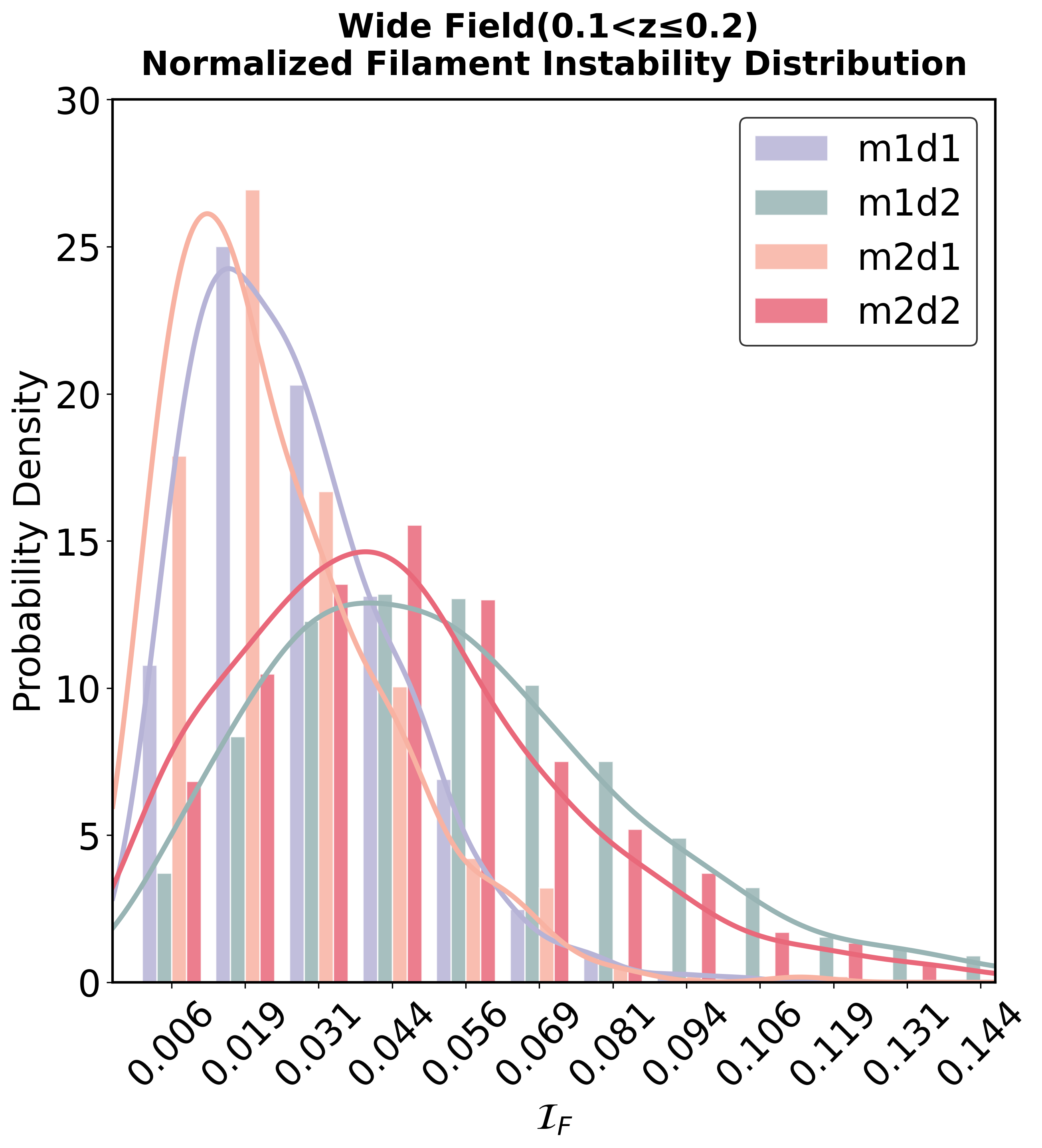}
\includegraphics[width=0.49\textwidth]{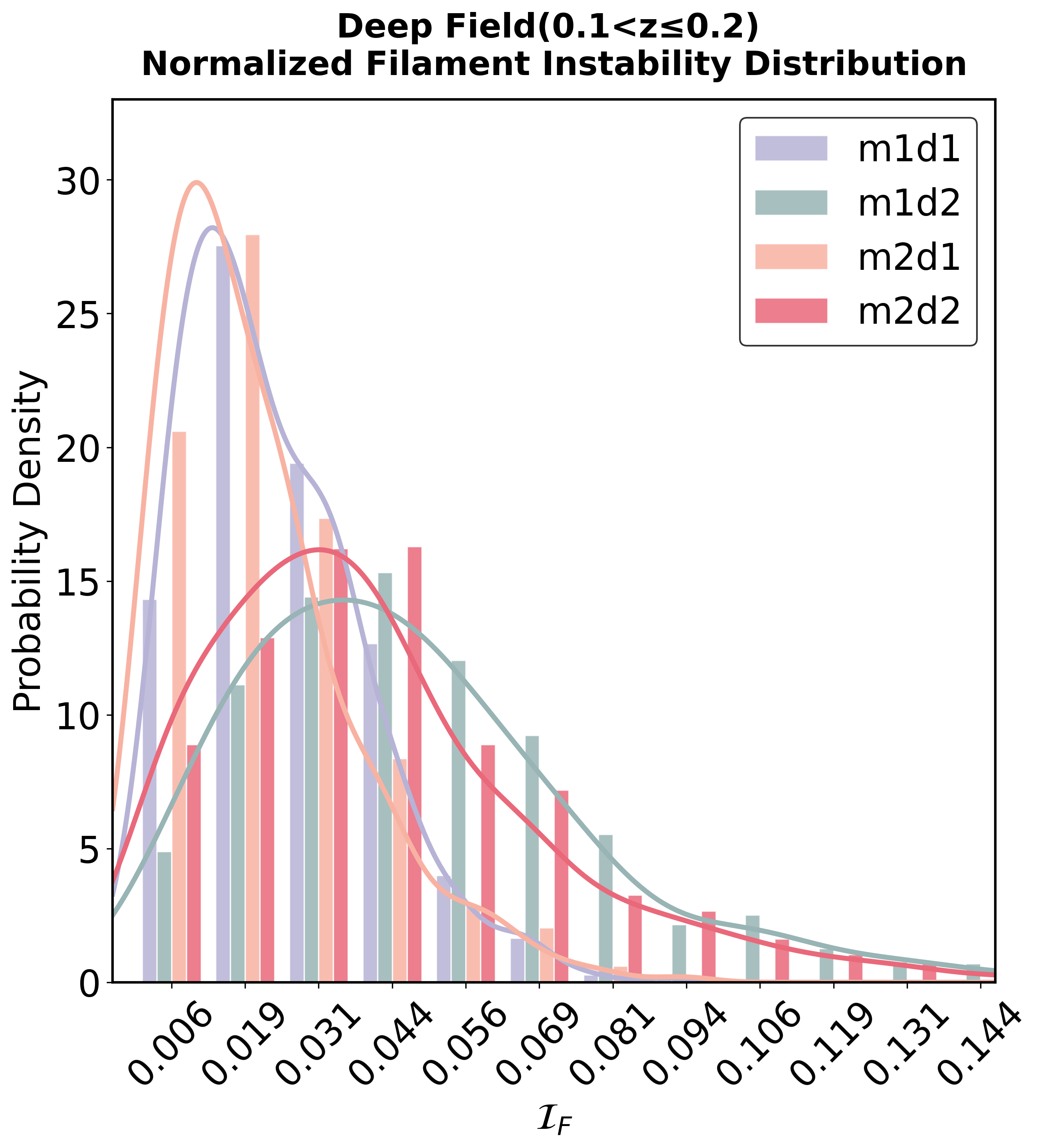}
\includegraphics[width=0.49\textwidth]{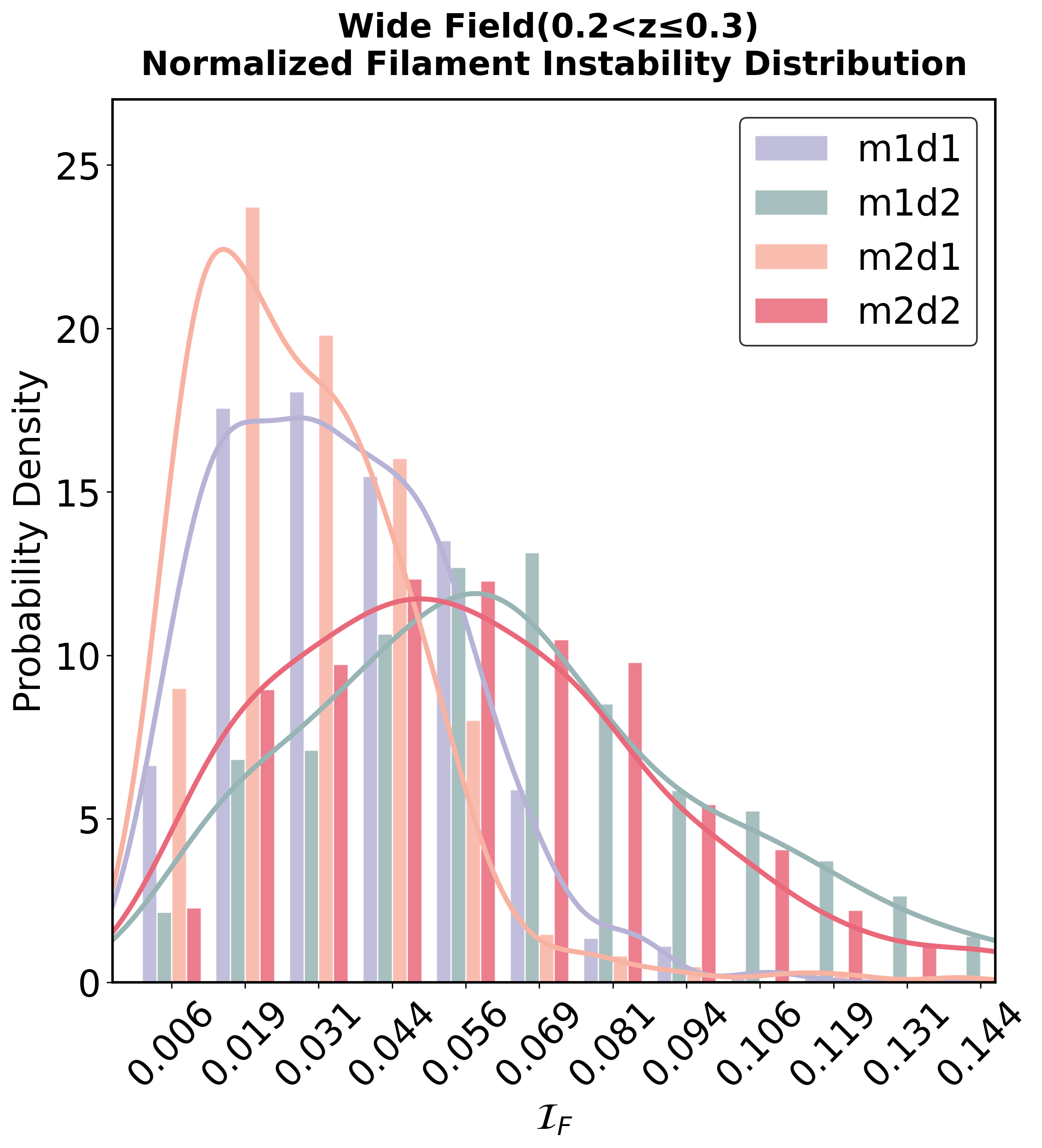}
\includegraphics[width=0.49\textwidth]{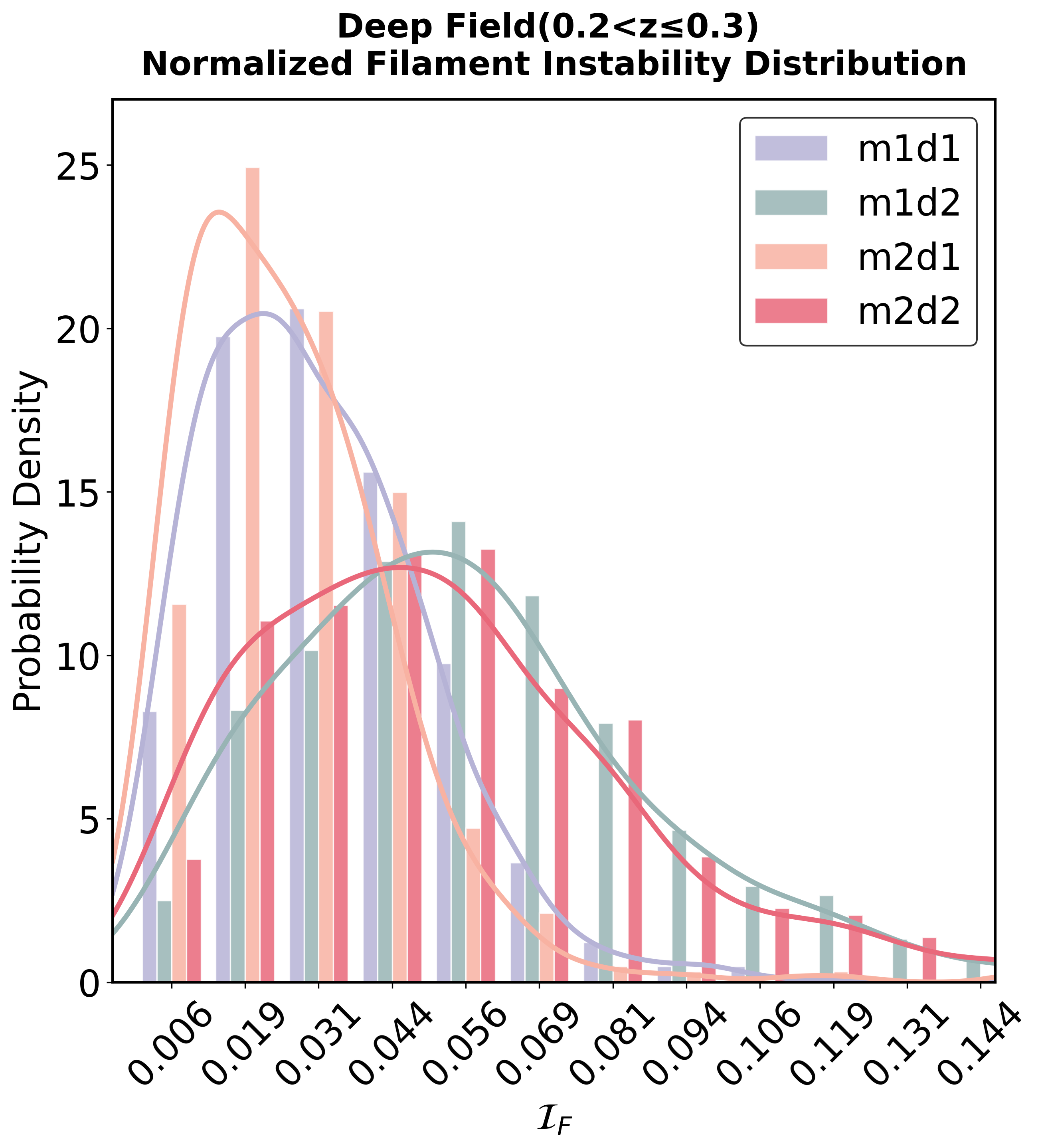}
\caption{Histograms of the filament-recovery instability parameter
$\mathcal{I}_F$ for the four subsamples defined in
Table~\ref{tab:bin}. The $z1$ ($0.1<z\leq0.2$) and $z2$
($0.2<z\leq0.3$) bins are presented in the top and bottom rows,
respectively, with the wide- and deep-field cases shown separately.
The overlaid KDE curves use bandwidths determined by Silverman's rule.
In the legends, $m1$ and $m2$ denote
$14.5<\log_{10}[M/(h^{-1}M_\odot)]\leq14.8$ and
$14.8<\log_{10}[M/(h^{-1}M_\odot)]\leq15.5$, respectively, while $d1$
and $d2$ denote $d_{\rm proj}<5\,h^{-1}\mathrm{Mpc}$ and
$5\leq d_{\rm proj}<10\,h^{-1}\mathrm{Mpc}$, respectively.
}

\label{fig:his_stability} 
\end{figure*}

\paragraph{Physical Properties}
The filament length $L$ is measured as the total distance along the traced path connecting the two halos. In addition, a dimensionless shape parameter $K$,
\begin{equation}
\mathcal{K} = \frac{L_{\rm filament}}{L_{\rm straight}},
\end{equation}
is introduced to quantify deviations of the filament from a straight connection between the two halo centers. To provide a physically meaningful normalization scale, we define a characteristic halo size as the mean virial radius of the halo pair,
\begin{equation}
\langle R_{200} \rangle = \frac{1}{2}\left(R_{200}^{(a)} + R_{200}^{(b)}\right).
\end{equation}
This characteristic radius is used solely as a reference length scale for comparing filament geometry across different halo pairs. When reference (``true'') filament paths are available, the robustness of filament recovery against noise is quantified using a filament recovery stability metric. For each noise realization, we measure the discrepancy by averaging the shortest distances from $\mathcal{P}_i$ to $\mathcal{P}_{\rm true}$ and from $\mathcal{P}_{\rm true}$ to $\mathcal{P}_i$, which is denoted by $dis(\mathcal{P}_i, \mathcal{P}_{\rm true})$. This distance is normalized by the characteristic halo scale $\langle R_{200} \rangle$ to define a dimensionless filament recovery instability parameter $\mathcal{I}_{F}$,
\begin{equation}
\mathcal{I}_{F} = \frac{\langle dis(\mathcal{P}_i, \mathcal{P}_{\rm true}) \rangle}{\langle R_{200} \rangle}.
\end{equation}
This metric directly quantifies the spatial reconstruction error relative to the halo scale, which is important for assessing the reliability of the recovered halo-to-halo connections and filament geometry. Smaller values of $\mathcal{I}_{F}$ correspond to lower instability, indicating that the recovered filament geometry is less sensitive to noise fluctuations and remains spatially consistent across different realizations.

\paragraph{Validation and Outputs}
The pipeline produces diagnostic visualizations and records all
filament properties, including filament length, shape parameter, and
recovery stability metrics, for quantitative comparison across halo
pairs. Representative visual examples are shown in
Fig.~\ref{fig:filament_z0102} and Fig.~\ref{fig:filament_z0203},
where the normalized path-probability maps and
projected density profiles derived from 50 noise realizations are
presented for both MIU$^{2}$Net and KS alongside the true results.
The visual comparison demonstrates that MIU$^{2}$Net accurately
recovers not only the main filamentary structures but also
low-$\kappa$ regions on small scales, which remain in good agreement
with the true maps in both wide- and deep-field cases.
Although the filament profiles recovered by KS broadly
follow the true profiles, their peak locations and small-scale features
are reproduced less accurately.

The corresponding numerical measurements are summarized in
Fig.~\ref{fig:his_stability}, Table~\ref{tab:stability}, and
Table~\ref{tab:shape}. Fig.~\ref{fig:his_stability}
shows that filament-recovery instability generally increases with halo
separation. As summarized in Table~\ref{tab:stability}, the
filament-recovery instability parameter $\mathcal{I}_F$ exhibits a
dependence on redshift, with lower mean values in the low-redshift bin
than in the high-redshift bin for both observing depths.
Specifically for MIU$^{2}$Net,
$\langle\mathcal{I}_F\rangle$ decreases from 0.041 to 0.037 in the
low-redshift bin and from 0.051 to 0.045 in the high-redshift bin when
moving from the wide- to the deep-field case, indicating a modest
improvement in filament-recovery stability under the lower-noise
condition. The corresponding KS measurements are
obtained under the same observing conditions but exhibit systematically
larger $\mathcal{I}_F$ values than MIU$^{2}$Net, indicating less stable
filament recovery.

Overall, the small magnitude of $\mathcal{I}_F$ for MIU$^{2}$Net in
both redshift bins indicates that filament-reconstruction errors remain
well below the characteristic halo scale, suggesting that the
identified filaments are robust to noise and accurately recovered
relative to $R_{200}$. However, $\mathcal{I}_F$ shows a mild increase
with halo separation, indicating that filament reconstruction becomes
progressively more challenging for widely separated systems. In addition, Table~\ref{tab:shape} reports the mean $\mathcal{K}$ in
each bin alongside the corresponding true value.

Both the true and reconstructed $\mathcal{K}$ values remain close to 1, indicating that the filaments in our sample are, on average, relatively straight. This is likely related in part to our selection of close halo pairs, for which the connecting structures are expected to be relatively straight. Importantly, both MIU$^{2}$Net and KS recover $\mathcal{K}$ well, with the reconstructed values broadly consistent with the corresponding true values in both redshift bins. For MIU$^{2}$Net, the wide- and deep-field values are broadly comparable and slightly larger than the true values, while the KS reconstructions are likewise close to the true $\mathcal{K}$ values, with only modest differences between the two field depths.

\begin{table*}
\centering

\caption{Filament Recovery Stability Parameter $\mathcal{I}_{F}$.}

\begin{tabular}{c|cccc}
\hline\hline
Sample
& MIU$^{2}$Net-Wide
& MIU$^{2}$Net-Deep
& KS-Wide
& KS-Deep \\
\hline

Pair(z1m1d1)
& $0.030_{+0.047}^{-0.013}$
& $0.027_{+0.042}^{-0.012}$
& $0.082_{+0.115}^{-0.050}$
& $0.075_{+0.107}^{-0.042}$ \\

Pair(z1m1d2)
& $0.058_{+0.092}^{-0.023}$
& $0.052_{+0.088}^{-0.015}$
& $0.144_{+0.202}^{-0.087}$
& $0.136_{+0.197}^{-0.075}$ \\

Pair(z1m2d1)
& $0.027_{+0.045}^{-0.009}$
& $0.024_{+0.040}^{-0.008}$
& $0.071_{+0.106}^{-0.037}$
& $0.064_{+0.100}^{-0.029}$ \\

Pair(z1m2d2)
& $0.049_{+0.079}^{-0.019}$
& $0.044_{+0.074}^{-0.015}$
& $0.122_{+0.181}^{-0.064}$
& $0.114_{+0.174}^{-0.053}$ \\

Pair(z2m1d1)
& $0.038_{+0.058}^{-0.018}$
& $0.034_{+0.053}^{-0.015}$
& $0.122_{+0.172}^{-0.073}$
& $0.110_{+0.158}^{-0.061}$ \\

Pair(z2m1d2)
& $0.070_{+0.111}^{-0.030}$
& $0.061_{+0.098}^{-0.024}$
& $0.206_{+0.289}^{-0.124}$
& $0.191_{+0.276}^{-0.105}$ \\

Pair(z2m2d1)
& $0.032_{+0.051}^{-0.013}$
& $0.030_{+0.049}^{-0.011}$
& $0.096_{+0.140}^{-0.051}$
& $0.086_{+0.130}^{-0.041}$ \\

Pair(z2m2d2)
& $0.062_{+0.084}^{-0.027}$
& $0.056_{+0.089}^{-0.023}$
& $0.168_{+0.241}^{-0.094}$
& $0.153_{+0.228}^{-0.078}$ \\

\hline\hline
\end{tabular}

\begin{tablenotes}[flushleft]
\item \textbf{Note}:
The subsample definitions follow Table~\ref{tab:bin}. The $z1$ and $z2$
bins correspond to $0.1<z\leq0.2$ and $0.2<z\leq0.3$, respectively.
The labels $m1$ and $m2$ denote
$14.5<\log_{10}[M/(h^{-1}M_\odot)]\leq14.8$ and
$14.8<\log_{10}[M/(h^{-1}M_\odot)]\leq15.5$, while $d1$ and $d2$
denote $d_{\rm proj}<5\,h^{-1}\mathrm{Mpc}$ and
$5\leq d_{\rm proj}<10\,h^{-1}\mathrm{Mpc}$, respectively.
\end{tablenotes}

\label{tab:stability}
\end{table*}

\begin{table*}
\centering

\caption{Filament Shape Parameter $\mathcal{K}$.}

\begin{tabular}{c|ccccc}
\hline\hline
Sample
& True
& MIU$^{2}$Net-Wide
& MIU$^{2}$Net-Deep
& KS-Wide
& KS-Deep \\
\hline

Pair(z1m1d1)
& 1.041
& 1.043
& 1.044
& 1.062
& 1.057 \\

Pair(z1m1d2)
& 1.039
& 1.043
& 1.043
& 1.053
& 1.049 \\

Pair(z1m2d1)
& 1.039
& 1.042
& 1.042
& 1.056
& 1.052 \\

Pair(z1m2d2)
& 1.039
& 1.042
& 1.043
& 1.050
& 1.048 \\

Pair(z2m1d1)
& 1.042
& 1.041
& 1.042
& 1.070
& 1.063 \\

Pair(z2m1d2)
& 1.038
& 1.041
& 1.041
& 1.059
& 1.054 \\

Pair(z2m2d1)
& 1.038
& 1.038
& 1.039
& 1.069
& 1.053 \\

Pair(z2m2d2)
& 1.041
& 1.043
& 1.043
& 1.056
& 1.053 \\

\hline\hline
\end{tabular}

\begin{tablenotes}[flushleft]
\item \textbf{Note}:
The subsample definitions are the same as those in
Table~\ref{tab:stability}. The complete definitions are provided in
Table~\ref{tab:bin}.
\end{tablenotes}

\label{tab:shape}
\end{table*}

\section{Conclusion}
\label{sec:conclusion}

The upcoming China Space Station Telescope (CSST) will provide an unprecedented opportunity to study the dark matter distribution through weak gravitational lensing. In this work, we investigate the feasibility of recovering complex dark matter structures in the CSST era, including filament-connected halo pairs and merging systems. Using CSST-like mock observations constructed from two cosmological simulations, we apply the deep-learning-based MIU$^{2}$Net reconstruction framework to assess the recoverability of these systems under both the planned wide-field and deep-field survey configurations.

Our results demonstrate that both the planned CSST wide-field and
deep-field surveys will deliver weak-lensing data of sufficient quality
to recover filamentary structures and complex dark-matter morphologies
with high fidelity. Halo centers are reconstructed with high positional
accuracy, with most offsets remaining within
$\sim 0.1\,h^{-1}\,\mathrm{Mpc}$ in both survey modes. Averaging over the wide- and deep-field cases, MIU$^{2}$Net achieves mean absolute axis-ratio biases of 0.044 and 0.063 for the low- and high-redshift bins, respectively.
Filamentary structures are also recovered robustly against noise by
MIU$^{2}$Net, as indicated by the low values of the filament-recovery
instability parameter $\mathcal{I}_F$, which measures the normalized
spatial discrepancy between the recovered and true filament paths
across noise realizations. For MIU$^{2}$Net, we find
$\langle\mathcal{I}_F\rangle \approx 0.041$ and $0.051$ in the low- and
high-redshift bins for the wide-field survey, improving to approximately
$0.037$ and $0.045$ for the deep-field samples.
The corresponding KS reconstructions are less stable.
The filament shape parameter $\mathcal{K}$ recovered by MIU$^{2}$Net,
which quantifies deviations from a straight halo-to-halo connection,
remains consistent with the true value at the percent level.
The filaments recovered by KS are slightly more curved. Overall, the increased source galaxy number density in the
deep-field survey primarily improves the stability of filament
reconstruction, while MIU$^{2}$Net maintains highly consistent
mass-morphology recovery between the two observing scenarios.


These results demonstrate that CSST weak-lensing observations, combined with machine-learning-based mass reconstruction, will enable detailed studies of filament-connected halo pairs, merging clusters, and the dark-matter distribution within the cosmic web. In particular, the ability to recover both cluster-scale structures and connecting filaments opens new opportunities for investigating cluster assembly, matter transport along filaments, and the role of environment in structure formation. Several limitations should be noted when interpreting these results.
Our current analysis adopts an idealized treatment of the PSF and does
not account for residual PSF-modeling uncertainties or
photometric-redshift uncertainties. CSST forecasts require the
residual multiplicative shear bias to be controlled to
$|m|\lesssim2.6\times10^{-3}$ \citep{yao2024csst}, while the expected
photometric-redshift scatter and catastrophic-outlier fraction are
$\sigma_z\sim0.02$ and $\sim3\%$, respectively \citep{Cao2018}. For the
test samples in our work, thin-lens approximation is adopted to highlight the
cluster and filament morphologies of interest. Foreground and background mass
concentrations are therefore not included in the current analysis. Moreover,
the present study focuses on relatively massive systems, and the reconstruction
performance is expected to degrade for lower-mass clusters because of their
weaker lensing signals. Although a complete treatment of these effects is
beyond the scope of this work, future training sets will include lower-mass
systems, full light-cone simulations, realistic PSF variations, and
photometric-redshift uncertainties.

\begin{acknowledgements}
We acknowledge the support by National Key R\&D Program of China No. 2022YFF0503403, the support of National Nature Science Foundation of China (No 11988101), the support from the Ministry of Science and Technology of China (No. 2020SKA0110100), CAS Project for Young Scientists in Basic Research (No. YSBR-062). X.K.L. acknowledges the support from NSFC grant No. 12173033, the grants from the China Manned Space Projects with No. CMS-CSST-2021-B01, the“Yunnan Key Laboratory of Survey Science” with project No. 202449CE340002 and a “Yunnan Provincial Top Team Project” with project No.202305AT350002. JY acknowledges the support from NSFC Grant No.12203084 and 12573006. We also acknowledge the support from the science research grants from the China Manned Space Project with No. CMS-CSST-2025-A03 and CMS-CSST-2025-A05.
\end{acknowledgements}

\bibliography{main} 

\begin{thebibliography}{}
\expandafter\ifx\csname natexlab\endcsname\relax\def\natexlab#1{#1}\fi
\providecommand{\url}[1]{\href{#1}{#1}}
\providecommand{\dodoi}[1]{doi:~\href{http://doi.org/#1}{\nolinkurl{#1}}}
\providecommand{\doeprint}[1]{\href{http://ascl.net/#1}{\nolinkurl{http://ascl.net/#1}}}
\providecommand{\doarXiv}[1]{\href{https://arxiv.org/abs/#1}{\nolinkurl{https://arxiv.org/abs/#1}}}

\bibitem[{Bartelmann \& Schneider(2001)}]{bartelmann2001weak}
Bartelmann, M., \& Schneider, P. 2001, Physics Reports, 340, 291,
  \dodoi{10.1016/S0370-1573(00)00082-X}

\bibitem[{{Blumenthal} {et~al.}(1984){Blumenthal}, {Faber}, {Primack}, \&
  {Rees}}]{1984Natur.311..517B}
{Blumenthal}, G.~R., {Faber}, S.~M., {Primack}, J.~R., \& {Rees}, M.~J. 1984,
  \nat, 311, 517, \dodoi{10.1038/311517a0}

\bibitem[{{Bond} {et~al.}(1996){Bond}, {Kofman}, \&
  {Pogosyan}}]{1996Natur.380..603B}
{Bond}, J.~R., {Kofman}, L., \& {Pogosyan}, D. 1996, \nat, 380, 603,
  \dodoi{10.1038/380603a0}

\bibitem[{Cao {et~al.}(2018)Cao, Gong, Meng, Xu, Chen, Guo, Li, Liu, Xue, Cao,
  Fu, Zhang, Wang, \& Zhan}]{Cao2018}
Cao, Y., Gong, Y., Meng, X.-M., {et~al.} 2018, Monthly Notices of the Royal
  Astronomical Society, 480, 2178, \dodoi{10.1093/mnras/sty1980}

\bibitem[{{Clowe} {et~al.}(2006){Clowe}, {Brada{\v{c}}}, {Gonzalez},
  {Markevitch}, {Randall}, {Jones}, \& {Zaritsky}}]{2006ApJ...648L.109C}
{Clowe}, D., {Brada{\v{c}}}, M., {Gonzalez}, A.~H., {et~al.} 2006, \apjl, 648,
  L109, \dodoi{10.1086/508162}

\bibitem[{{Colberg} {et~al.}(2005){Colberg}, {Krughoff}, \&
  {Connolly}}]{2005MNRAS.359..272C}
{Colberg}, J.~M., {Krughoff}, K.~S., \& {Connolly}, A.~J. 2005, \mnras, 359,
  272, \dodoi{10.1111/j.1365-2966.2005.08897.x}

\bibitem[{{CSST Collaboration} {et~al.}(2026){CSST Collaboration}, {Gong},
  {Miao}, {Zhan}, {Li}, {Shangguan}, {Li}, {Liu}, {Chen}, {Yuan}, {Zhou},
  {Liu}, {Yu}, {Ji}, {Qi}, {Liu}, {Dai}, {Wang}, {Zheng}, {Hao}, {Dou}, {Ao},
  {Lin}, {Zhang}, {Wang}, {Sun}, {Li}, {Li}, {Xu}, {Li}, {Li}, {Wu}, {Zhang},
  {Wang}, {Bai}, {Cai}, {Cai}, {Cao}, {Chan}, {Chang}, {Chen}, {Chen}, {Chen},
  {Chen}, {Cui}, {Dong}, {Du}, {Duan}, {Fan}, {Fan}, {Fan}, {Fan}, {Fang},
  {Fu}, {Fu}, {Fu}, {Gao}, {Gu}, {Gu}, {Guo}, {Han}, {Hu}, {Huang}, {Ho},
  {Jiang}, {Jiang}, {Jing}, {Kang}, {Kong}, {Li}, {Li}, {Li}, {Li}, {Li}, {Li},
  {Liao}, {Lin}, {Liu}, {Liu}, {Liu}, {Liu}, {Mao}, {Mao}, {Meng}, {Pang},
  {Peng}, {Peng}, {Shan}, {Shen}, {Shen}, {Shen}, {Shi}, {Shi}, {Tan}, {Tian},
  {Wang}, {Wang}, {Wang}, {Wang}, {Wu}, {Wu}, {Wu}, {Xu}, {Xue}, {Xue}, {Yang},
  {Yang}, {Yao}, {Yuan}, {Yuan}, {Zhang}, {Zhang}, {Zhang}, {Zhang}, {Zhang},
  {Zhao}, {Zhao}, {Zhong}, {Zhong}, {Zhou}, {Zhu}, \&
  {Zu}}]{2026SCPMA..6939501C}
{CSST Collaboration}, {Gong}, Y., {Miao}, H., {et~al.} 2026, Science China
  Physics, Mechanics, and Astronomy, 69, 239501,
  \dodoi{10.1007/s11433-025-2809-0}

\bibitem[{{Davis} {et~al.}(1985){Davis}, {Efstathiou}, {Frenk}, \&
  {White}}]{1985ApJ...292..371D}
{Davis}, M., {Efstathiou}, G., {Frenk}, C.~S., \& {White}, S.~D.~M. 1985, \apj,
  292, 371, \dodoi{10.1086/163168}

\bibitem[{Dietrich {et~al.}(2012)Dietrich, Werner, Clowe, Finoguenov, Kitching,
  Miller, \& Simionescu}]{dietrich2012filament}
Dietrich, J.~P., Werner, N., Clowe, D., {et~al.} 2012, Nature, 487, 202

\bibitem[{{Forero-Romero} {et~al.}(2014){Forero-Romero}, {Contreras}, \&
  {Padilla}}]{2014MNRAS.443.1090F}
{Forero-Romero}, J.~E., {Contreras}, S., \& {Padilla}, N. 2014, \mnras, 443,
  1090, \dodoi{10.1093/mnras/stu1150}

\bibitem[{{Hamana} {et~al.}(2004){Hamana}, {Takada}, \&
  {Yoshida}}]{2004MNRAS.350..893H}
{Hamana}, T., {Takada}, M., \& {Yoshida}, N. 2004, \mnras, 350, 893,
  \dodoi{10.1111/j.1365-2966.2004.07691.x}

\bibitem[{{Han} {et~al.}(2026){Han}, {Zhao}, {Chen}, {Li}, {Li}, {Liu}, {Chen},
  \& {Yu}}]{2026RAA....26g5003H}
{Han}, W.~G., {Zhao}, A., {Chen}, X., {et~al.} 2026, Research in Astronomy and
  Astrophysics, 26, 075003, \dodoi{10.1088/1674-4527/ae4d1e}

\bibitem[{{Hellwing} {et~al.}(2021){Hellwing}, {Cautun}, {van de Weygaert}, \&
  {Jones}}]{2021PhRvD.103f3517H}
{Hellwing}, W.~A., {Cautun}, M., {van de Weygaert}, R., \& {Jones}, B.~T. 2021,
  \prd, 103, 063517, \dodoi{10.1103/PhysRevD.103.063517}

\bibitem[{Hennawi \& Spergel(2005)}]{Hennawi_2005}
Hennawi, J.~F., \& Spergel, D.~N. 2005, The Astrophysical Journal, 624,
  59–79, \dodoi{10.1086/428749}

\bibitem[{{Hilbert} {et~al.}(2009){Hilbert}, {Hartlap}, {White}, \&
  {Schneider}}]{2009A&A...499...31H}
{Hilbert}, S., {Hartlap}, J., {White}, S.~D.~M., \& {Schneider}, P. 2009, \aap,
  499, 31, \dodoi{10.1051/0004-6361/200811054}

\bibitem[{{Hong} {et~al.}(2021){Hong}, {Park}, {Jee}, {Bak}, \&
  {Cha}}]{2021ApJ...923..266H}
{Hong}, S.~E., {Park}, S., {Jee}, M.~J., {Bak}, D., \& {Cha}, S. 2021, \apj,
  923, 266, \dodoi{10.3847/1538-4357/ac3090}

\bibitem[{{Jeffrey} {et~al.}(2020){Jeffrey}, {Lanusse}, {Lahav}, \&
  {Starck}}]{2020MNRAS.492.5023J}
{Jeffrey}, N., {Lanusse}, F., {Lahav}, O., \& {Starck}, J.-L. 2020, \mnras,
  492, 5023, \dodoi{10.1093/mnras/staa127}

\bibitem[{Jeffrey {et~al.}(2018)Jeffrey, Abdalla, Lahav, Lanusse, Starck,
  Leonard, Kirk, Chang, Baxter, Kacprzak, Seitz, Vikram, Whiteway, Abbott,
  Allam, Avila, Bertin, Brooks, Carnero Rosell, Carrasco Kind, Carretero,
  Castander, Crocce, Cunha, D’Andrea, da Costa, Davis, De Vicente, Desai,
  Doel, Eifler, Evrard, Flaugher, Fosalba, Frieman, García-Bellido, Gerdes,
  Gruen, Gruendl, Gschwend, Gutierrez, Hartley, Honscheid, Hoyle, James,
  Jarvis, Kuehn, Lima, Lin, March, Melchior, Menanteau, Miquel, Plazas, Reil,
  Roodman, Sanchez, Scarpine, Schubnell, Sevilla-Noarbe, Smith, Soares-Santos,
  Sobreira, Suchyta, Swanson, Tarle, Thomas, \& Walker}]{Jeffrey_2018}
Jeffrey, N., Abdalla, F.~B., Lahav, O., {et~al.} 2018, Monthly Notices of the
  Royal Astronomical Society, 479, 2871–2888, \dodoi{10.1093/mnras/sty1252}

\bibitem[{{Kaiser} \& {Squires}(1993)}]{1993ApJ...404..441K}
{Kaiser}, N., \& {Squires}, G. 1993, \apj, 404, 441, \dodoi{10.1086/172297}

\bibitem[{{Leauthaud} {et~al.}(2007){Leauthaud}, {Massey}, {Kneib}, {Rhodes},
  {Johnston}, {Capak}, {Heymans}, {Ellis}, {Koekemoer}, {Le F{\`e}vre},
  {Mellier}, {R{\'e}fr{\'e}gier}, {Robin}, {Scoville}, {Tasca}, {Taylor}, \&
  {Van Waerbeke}}]{2007ApJS..172..219L}
{Leauthaud}, A., {Massey}, R., {Kneib}, J.-P., {et~al.} 2007, \apjs, 172, 219,
  \dodoi{10.1086/516598}

\bibitem[{Lewis {et~al.}(2000)Lewis, Challinor, \& Lasenby}]{Lewis:1999bs}
Lewis, A., Challinor, A., \& Lasenby, A. 2000, Astrophys. J., 538, 473,
  \dodoi{10.1086/309179}

\bibitem[{Liu {et~al.}(2023)Liu, Meng, Er, Fan, Kilbinger, Li, Li, Schrabback,
  Scognamiglio, Shan, {et~al.}}]{liu2023potential}
Liu, D., Meng, X., Er, X., {et~al.} 2023, Astronomy \& Astrophysics, 669, A128

\bibitem[{Liu \& Madhavacheril(2019)}]{Liu_2019}
Liu, J., \& Madhavacheril, M.~S. 2019, Physical Review D, 99,
  \dodoi{10.1103/physrevd.99.083508}

\bibitem[{{Liu} {et~al.}(2015){Liu}, {Pan}, {Li}, {Shan}, {Wang}, {Fu}, {Fan},
  {Kneib}, {Leauthaud}, {Van Waerbeke}, {Makler}, {Moraes}, {Erben}, \&
  {Charbonnier}}]{2015MNRAS.450.2888L}
{Liu}, X., {Pan}, C., {Li}, R., {et~al.} 2015, \mnras, 450, 2888,
  \dodoi{10.1093/mnras/stv784}

\bibitem[{{Mandelbaum}(2018)}]{2018ARA&A..56..393M}
{Mandelbaum}, R. 2018, \araa, 56, 393,
  \dodoi{10.1146/annurev-astro-081817-051928}

\bibitem[{Markevitch {et~al.}(2004)Markevitch, Gonzalez, Clowe, Vikhlinin,
  Forman, Jones, Murray, \& Tucker}]{Markevitch_2004}
Markevitch, M., Gonzalez, A.~H., Clowe, D., {et~al.} 2004, The Astrophysical
  Journal, 606, 819–824, \dodoi{10.1086/383178}

\bibitem[{Maturi {et~al.}(2005)Maturi, Meneghetti, Bartelmann, Dolag, \&
  Moscardini}]{Maturi_2005}
Maturi, M., Meneghetti, M., Bartelmann, M., Dolag, K., \& Moscardini, L. 2005,
  Astronomy \& Astrophysics, 442, 851–860, \dodoi{10.1051/0004-6361:20042600}

\bibitem[{{Peebles}(1980)}]{1980lssu.book.....P}
{Peebles}, P.~J.~E. 1980, {The large-scale structure of the universe}

\bibitem[{{Planck Collaboration} {et~al.}(2020){Planck Collaboration},
  {Aghanim}, {Akrami}, {Ashdown}, {Aumont}, {Baccigalupi}, {Ballardini},
  {Banday}, {Barreiro}, {Bartolo}, {Basak}, {Battye}, {Benabed}, {Bernard},
  {Bersanelli}, {Bielewicz}, {Bock}, {Bond}, {Borrill}, {Bouchet}, {Boulanger},
  {Bucher}, {Burigana}, {Butler}, {Calabrese}, {Cardoso}, {Carron},
  {Challinor}, {Chiang}, {Chluba}, {Colombo}, {Combet}, {Contreras}, {Crill},
  {Cuttaia}, {de Bernardis}, {de Zotti}, {Delabrouille}, {Delouis}, {Di
  Valentino}, {Diego}, {Dor{\'e}}, {Douspis}, {Ducout}, {Dupac}, {Dusini},
  {Efstathiou}, {Elsner}, {En{\ss}lin}, {Eriksen}, {Fantaye}, {Farhang},
  {Fergusson}, {Fernandez-Cobos}, {Finelli}, {Forastieri}, {Frailis},
  {Fraisse}, {Franceschi}, {Frolov}, {Galeotta}, {Galli}, {Ganga},
  {G{\'e}nova-Santos}, {Gerbino}, {Ghosh}, {Gonz{\'a}lez-Nuevo}, {G{\'o}rski},
  {Gratton}, {Gruppuso}, {Gudmundsson}, {Hamann}, {Handley}, {Hansen},
  {Herranz}, {Hildebrandt}, {Hivon}, {Huang}, {Jaffe}, {Jones}, {Karakci},
  {Keih{\"a}nen}, {Keskitalo}, {Kiiveri}, {Kim}, {Kisner}, {Knox},
  {Krachmalnicoff}, {Kunz}, {Kurki-Suonio}, {Lagache}, {Lamarre}, {Lasenby},
  {Lattanzi}, {Lawrence}, {Le Jeune}, {Lemos}, {Lesgourgues}, {Levrier},
  {Lewis}, {Liguori}, {Lilje}, {Lilley}, {Lindholm}, {L{\'o}pez-Caniego},
  {Lubin}, {Ma}, {Mac{\'\i}as-P{\'e}rez}, {Maggio}, {Maino}, {Mandolesi},
  {Mangilli}, {Marcos-Caballero}, {Maris}, {Martin}, {Martinelli},
  {Mart{\'\i}nez-Gonz{\'a}lez}, {Matarrese}, {Mauri}, {McEwen}, {Meinhold},
  {Melchiorri}, {Mennella}, {Migliaccio}, {Millea}, {Mitra},
  {Miville-Desch{\^e}nes}, {Molinari}, {Montier}, {Morgante}, {Moss}, {Natoli},
  {N{\o}rgaard-Nielsen}, {Pagano}, {Paoletti}, {Partridge}, {Patanchon},
  {Peiris}, {Perrotta}, {Pettorino}, {Piacentini}, {Polastri}, {Polenta},
  {Puget}, {Rachen}, {Reinecke}, {Remazeilles}, {Renzi}, {Rocha}, {Rosset},
  {Roudier}, {Rubi{\~n}o-Mart{\'\i}n}, {Ruiz-Granados}, {Salvati}, {Sandri},
  {Savelainen}, {Scott}, {Shellard}, {Sirignano}, {Sirri}, {Spencer},
  {Sunyaev}, {Suur-Uski}, {Tauber}, {Tavagnacco}, {Tenti}, {Toffolatti},
  {Tomasi}, {Trombetti}, {Valenziano}, {Valiviita}, {Van Tent}, {Vibert},
  {Vielva}, {Villa}, {Vittorio}, {Wandelt}, {Wehus}, {White}, {White},
  {Zacchei}, \& {Zonca}}]{2020A&A...641A...6P}
{Planck Collaboration}, {Aghanim}, N., {Akrami}, Y., {et~al.} 2020, \aap, 641,
  A6, \dodoi{10.1051/0004-6361/201833910}

\bibitem[{Qin {et~al.}(2020)Qin, Zhang, Huang, Dehghan, Zaiane, \&
  Jagersand}]{qin2020u2}
Qin, X., Zhang, Z., Huang, C., {et~al.} 2020, Pattern recognition, 106, 107404

\bibitem[{{Randall} {et~al.}(2008){Randall}, {Markevitch}, {Clowe}, {Gonzalez},
  \& {Brada{\v{c}}}}]{2008ApJ...679.1173R}
{Randall}, S.~W., {Markevitch}, M., {Clowe}, D., {Gonzalez}, A.~H., \&
  {Brada{\v{c}}}, M. 2008, \apj, 679, 1173, \dodoi{10.1086/587859}

\bibitem[{{Schneider} {et~al.}(2017){Schneider}, {Ng}, {Dawson}, {Marshall},
  {Meyers}, \& {Bard}}]{2017ApJ...839...25S}
{Schneider}, M.~D., {Ng}, K.~Y., {Dawson}, W.~A., {et~al.} 2017, \apj, 839, 25,
  \dodoi{10.3847/1538-4357/839/1/25}

\bibitem[{{Shapiro}(2009)}]{2009ApJ...696..775S}
{Shapiro}, C. 2009, \apj, 696, 775, \dodoi{10.1088/0004-637X/696/1/775}

\bibitem[{{Shirasaki} {et~al.}(2019){Shirasaki}, {Yoshida}, \&
  {Ikeda}}]{2019PhRvD.100d3527S}
{Shirasaki}, M., {Yoshida}, N., \& {Ikeda}, S. 2019, \prd, 100, 043527,
  \dodoi{10.1103/PhysRevD.100.043527}

\bibitem[{{Springel} {et~al.}(2005){Springel}, {White}, {Jenkins}, {Frenk},
  {Yoshida}, {Gao}, {Navarro}, {Thacker}, {Croton}, {Helly}, {Peacock}, {Cole},
  {Thomas}, {Couchman}, {Evrard}, {Colberg}, \& {Pearce}}]{2005Natur.435..629S}
{Springel}, V., {White}, S. D.~M., {Jenkins}, A., {et~al.} 2005, \nat, 435,
  629, \dodoi{10.1038/nature03597}

\bibitem[{Starck {et~al.}(2006)Starck, Pires, \& Réfrégier}]{Starck_2006}
Starck, J.-L., Pires, S., \& Réfrégier, A. 2006, Astronomy \& Astrophysics,
  451, 1139–1150, \dodoi{10.1051/0004-6361:20052997}

\bibitem[{Tam {et~al.}(2020)Tam, Massey, Jauzac, \& Robertson}]{Tam_2020}
Tam, S.-I., Massey, R., Jauzac, M., \& Robertson, A. 2020, Monthly Notices of
  the Royal Astronomical Society, 496, 3973–3990,
  \dodoi{10.1093/mnras/staa1756}

\bibitem[{Wang {et~al.}(2026)Wang, Li, Gao, Guo, Angulo, Cha, Cole, Frenk,
  HyeongHan, Li, Pei, Shan, Wang, \& White}]{Wang_2026}
Wang, Q., Li, M., Gao, L., {et~al.} 2026, Monthly Notices of the Royal
  Astronomical Society, \dodoi{10.1093/mnras/stag540}

\bibitem[{Yao {et~al.}(2024)Yao, Shan, Li, Xu, Fan, Liu, Zhang, Yu, Wei, Hu,
  {et~al.}}]{yao2024csst}
Yao, J., Shan, H., Li, R., {et~al.} 2024, Monthly Notices of the Royal
  Astronomical Society, 527, 5206

\end{thebibliography}
\bibliographystyle{aasjournal}

\end{document}